\documentclass[11pt]{article}

\pdfoutput=1

\usepackage[T1]{fontenc}
\usepackage[latin9]{inputenc}
\usepackage[a4paper]{geometry}
\usepackage[active]{srcltx}
\usepackage{amsmath}
\usepackage{amssymb}
\usepackage{esint}

\makeatletter
\usepackage{textcomp}

\pdfoutput=1 % if your are submitting a pdflatex (i.e. if you have
\usepackage{jheppub}% for details on the use of the package, please
\usepackage{etoolbox}% http://ctan.org/pkg/etoolbox
    
    \patchcmd{\maketitle}{\@fpheader}{}{}{}

\usepackage{amsfonts}

\usepackage{bbm}

\title{\textbf{Asymptotic structure of $\mathcal{N}=2$ supergravity in 3D: 
extended super-BMS$_3$ and nonlinear energy bounds}}

\author[]{Oscar Fuentealba,}
\author[]{Javier Matulich,}
\author[]{Ricardo Troncoso}

\affiliation[]{Centro de Estudios Cient\'{i}ficos (CECs), Av. Arturo Prat 514, Valdivia,
Chile.}

\emailAdd{fuentealba@cecs.cl}
\emailAdd{matulich@cecs.cl}
\emailAdd{troncoso@cecs.cl}

\preprint{CECS-PHY-16/06}

\abstract{The asymptotically flat structure of $\mathcal{N}=(2,0)$ supergravity 
in three spacetime dimensions is explored. The asymptotic symmetries are found 
to be spanned by an extension of the super-BMS$_3$ algebra, endowed with two independent 
affine $\hat{u}(1)$ currents of electric and magnetic type. These currents are associated
 to $U(1)$ fields being even and odd under parity, respectively. Remarkably, although
 the $U(1)$ fields do not generate a backreaction on the metric, they provide nontrivial 
Sugawara-like contributions to the BMS$_3$ generators, and hence to the energy and the 
angular momentum. Consequently, the entropy of flat cosmological spacetimes endowed with
 $U(1)$ fields acquires a nontrivial dependence on the zero modes of the $\hat{u}(1)$ charges.
 If the spin structure is odd, the ground state corresponds to Minkowski spacetime, 
and although the anticommutator of the canonical supercharges is linear in the energy 
and in the electric-like $\hat{u}(1)$ charge, the energy becomes bounded from below by 
the energy of the ground state shifted by the square of the electric-like $\hat{u}(1)$ 
charge. If the spin structure is even, the same bound for the energy generically holds, 
unless the absolute value of the electric-like charge is less than minus the mass of 
Minkowski spacetime in vacuum, so that the energy has to be nonnegative. The explicit form of the 
global and asymptotic Killing spinors is found for a wide class of configurations that 
fulfills our boundary conditions, and they exist precisely when the corresponding bounds are saturated. It is also shown that the spectra with periodic or antiperiodic boundary conditions for the fermionic fields are related by spectral flow, in a similar way as it occurs for the $\mathcal{N}=2$ super-Virasoro algebra. Indeed, our supersymmetric extension of BMS$_3$ can be recovered from the Inönü-Wigner contraction of the superconformal algebra with $\mathcal{N}=(2,2)$, once the fermionic generators of the right copy are truncated. }

\makeatother

\begin{document}
\maketitle \flushbottom

\newpage{}

\section{Introduction}

The symmetries of asymptotically flat spacetimes at null infinity
were proposed to be spanned by the BMS algebra since long ago \cite{Bondi:1962px},
\cite{Sachs:1962wk}. More recently, this analysis has been further
developed and expanded in \cite{Barnich:2009se}, \cite{Barnich:2010eb},
\cite{Barnich:2011ct}, \cite{Barnich:2011mi}, \cite{Strominger:2013jfa},
\cite{Cachazo:2014fwa}, \cite{Kapec:2014opa}, \cite{Campiglia:2015yka},
and it has led to the proposal of \cite{Hawking:2016msc}, \cite{Hawking:2016sgy},
which might be promising in order to resolve the information loss
paradox \cite{Hawking:2015qqa}. Nonetheless, some open issues still
remain to be suitably understood in the four-dimensional case (see
e.g., \cite{Bousso:2017dny}), which naturally motivates one to explore
them in a simplified setup, as it is the case of General Relativity
in three spacetime dimensions. As shown in \cite{Ashtekar:1996cd},
\cite{Barnich:2006av}, \cite{Barnich:2010eb}, the three-dimensional
version of the BMS algebra (BMS$_{3}$) describes the asymptotically
flat symmetries. The BMS$_{3}$ algebra turns out to be isomorphic
to the Galilean conformal algebra in two dimensions, and it has been
shown to be relevant in the context of flat holography \cite{Bagchi:2009my},
\cite{Bagchi:2010zz}, \cite{Bagchi:2016geg}, \cite{Bagchi:2017cpu},
as well as for the tensionless limit of string theory \cite{Bagchi:2013bga},
\cite{Casali:2016atr}, \cite{Bagchi:2016yyf} (see also \cite{Mandal:2016lsa}).
It is also worth pointing out that the generators of the BMS$_{3}$
algebra can be seen to emerge in a unique way through a twisted Sugawara-like
construction made out from composite operators of affine currents
describing the asymptotic symmetries of the ``soft hairy'' type of
boundary conditions recently discussed in \cite{Afshar:2016wfy},
\cite{Grumiller:2016kcp}, \cite{Afshar:2016kjj}. Similar results
along these lines have also been found in the context of near horizon
(twisted) warped conformal symmetry algebras in \cite{Donnay:2015abr},
\cite{Afshar:2015wjm}, \cite{Donnay:2016ejv}.

Besides, in the context of $\mathcal{N}=1$ supergravity in three
spacetime dimensions \cite{Deser:1982sw}, \cite{S. Deser. 1984},
\cite{Marcus:1983hb}, the minimal supersymmetric extension of BMS$_{3}$
has been shown to arise from a suitable set of asymptotically flat
boundary conditions \cite{Barnich:2014cwa}, which are not necessarily
given at null infinity. The superalgebra turns out to be isomorphic
to the supersymmetric extension of the two-dimensional Galilean conformal
algebra in \cite{Bagchi:2009ke}, \cite{Mandal:2010gx} (see also
\cite{Banerjee:2015kcx}), which was found from a non-relativistic
limit of the superconformal algebra, and hence their generators do
not possess the same physical interpretation. 

The extension to the case of $\mathcal{N}=(1,1)$ has also been recently
explored in \cite{Lodato:2016alv}, where it was shown that two inequivalent
possibilities can be recovered from different flat limiting processes
of $\mathcal{N}=(1,1)$ AdS$_{3}$ supergravity \cite{Achucarro:1987vz}.
The homogeneous or ``democratic'' possibility corresponds to the straightforward
extension of the case with $\mathcal{N}=1$, which agrees with the
results found in \cite{Casali:2016atr}, \cite{Bagchi:2016yyf}, \cite{Banerjee:2016nio}
in the context of Galilean superconformal algebras. In the ``despotic''
possibility, the algebra becomes isomorphic to the inhomogeneous Galilean
superconformal algebra \cite{Mandal:2010gx}, \cite{Bagchi:2016yyf}.

In the next section we make a brief revision of the $\mathcal{N}=(2,0)$
Poincaré supergravity theory constructed out in \cite{Howe:1995zm}.
We show that demanding the action to be parity-invariant implies that
the $U(1)$ field that is minimally coupled to the complexified gravitino
is even under parity, while the remaining $U(1)$ field has to be
odd. In section \ref{Asympt} we propose a set of boundary conditions
that includes a generic choice of Lagrange multipliers, which is strictly
necessary in order to accommodate solutions of physical interest.
The asymptotic symmetries are shown to be spanned by a supersymmetric
extension of the BMS$_{3}$ algebra with $\mathcal{N}=(2,0)$, endowed
with two independent affine $\hat{u}(1)$ currents of electric and
magnetic type. Specifically, the nonvanishing anticommutator of the
complexified fermionic generators acquires a central extension and
depends on the supertranslations as well as on the electric-like affine
$\hat{u}(1)$ current. It is also shown that our supersymmetric extension
of BMS$_{3}$ can be recovered from the Inönü-Wigner contraction of
the superconformal algebra with $\mathcal{N}=(2,2)$, once the fermionic
generators of the right copy are truncated. In section \ref{Bounds}
we show that for fermionic fields that fulfill antiperiodic boundary
conditions, the ground state is given by Minkowski spacetime, possibly
endowed with $\hat{u}(1)$ charges of electric type. Remarkably, although
the anticommutator of the supercharges is linear in the energy and
in the electric-like $\hat{u}(1)$ charge, the energy becomes bounded
from below by the energy of Minkowski spacetime in vacuum, shifted
by the square of the electric-like $\hat{u}(1)$ charge. If the spin
structure is even, the same bound for the energy generically holds,
unless the absolute value of the electric-like charge is less than
minus the mass of Minkowski spacetime in vacuum, so that the energy
has to be nonnegative.

Bosonic configurations that fulfill our boundary conditions are revisited
in section \ref{Configurations}, where we pay special attention to
the conditions that ensure their regularity in terms of gauge fields.
We also discuss them in the metric formalism, and carry out a thorough
analysis of the thermodynamic properties of cosmological spacetimes
endowed with $U(1)$ fields. The presence of $U(1)$ fields of electric
and magnetic type also unveils some remarkable properties of Minkowski
spacetime, as well as locally flat configurations with conical defects
or surpluses. In section \ref{Bosonic} we focus in the analysis of
bosonic solutions possessing unbroken supersymmetries that saturate
the corresponding energy bounds, and we also find the explicit form
of the (asymptotic) Killing spinors. Section \ref{Spectral} is devoted
to show that the spectra with periodic or antiperiodic boundary conditions
for the fermionic fields are related by spectral flow, in a similar
way as it occurs for the super-Virasoro algebra $\mathcal{N}=2$ \cite{Schwimmer:1986mf}.
We conclude with some comments about the extension of our results
in section \ref{Extensions}. Our conventions are discussed in appendix
\ref{Conventions}.

Note added: while this manuscript was in the process of typesetting,
ref. \cite{Banerjee:2017gzj} was posted in the arxiv, which possesses
some overlap with particular cases of our results. 

\section{$\mathcal{N}=\left(2,0\right)$ Poincaré supergravity in three spacetime
dimensions}

As shown in \cite{Howe:1995zm}, $\mathcal{N}=\left(2,0\right)$ Poincaré
supergravity in three spacetime dimensions can be formulated as a
Chern-Simons theory for a suitable extension of the super-Poincaré
group. The algebra has to be endowed with two additional bosonic $U(1)$
generators that respectively correspond to an automorphism and a central
charge, so that it admits a non-degenerate invariant bilinear form.
The nonvanishing (anti-)commutators read
\begin{gather}
[J_{a},J_{b}]=\epsilon_{abc}J^{c}\quad,\quad[J_{a},P_{b}]=\epsilon_{abc}P^{c}\,,\nonumber \\{}
[J_{a},Q_{\alpha}^{I}]=\frac{1}{2}\left(\Gamma_{a}\right)_{\,\,\,\alpha}^{\beta}Q_{\beta}^{I}\quad,\quad[Q_{\alpha}^{I},T]=\epsilon^{IJ}Q_{\alpha}^{J}\,,\label{eq:Cov-Algebra}\\
\{Q_{\alpha}^{I},Q_{\beta}^{J}\}=-\frac{1}{2}\delta^{IJ}\left(C\Gamma^{a}\right)_{\alpha\beta}P_{a}+C_{\alpha\beta}\epsilon^{IJ}Z\,,\nonumber 
\end{gather}
where $C_{\alpha\beta}$ and $\left(\Gamma_{a}\right)_{\,\,\,\beta}^{\alpha}$
stand for the charge conjugation and Dirac matrices, respectively
(for our conventions see appendix \ref{Conventions}).

The existence of a nontrivial Casimir operator, given by 
\begin{equation}
I=2J^{a}P_{a}-Q_{\alpha}^{I}C^{\alpha\beta}Q_{\beta}^{I}-2TZ\,,\label{eq:Casimir}
\end{equation}
allows to define an invariant bilinear form whose nonvanishing components
read
\begin{equation}
\langle J_{a},P_{b}\rangle=\eta_{ab}\quad,\quad\langle Q_{\alpha}^{I},Q_{\beta}^{J}\rangle=C_{\alpha\beta}\delta^{IJ}\quad,\quad\langle T,Z\rangle=-1\,.\label{eq:bracketCov}
\end{equation}
It can be seen that this nondegenerate metric turns out to be the
unique one that leads to a parity even action, once suitable parity
properties of the fields are taken into account (see below). 

The entire field content can be arranged within a single connection
for the gauge supergroup
\begin{equation}
A=e^{a}P_{a}+\omega^{a}J_{a}+\psi_{I}^{\alpha}Q_{\alpha}^{I}+B\,T+C\,Z\,,\label{eq:A}
\end{equation}
so that apart from the dreibein $e^{a}$, the (dualized) spin connection
$\omega^{a}$, and the $\mathcal{N}=2$ gravitini $\psi_{I}^{\alpha}$,
the theory possesses two additional $U(1)$ fields, given by $B$
and $C$, respectively being even and odd under parity. 

The supergravity theory can then be described by a Chern-Simons action
\begin{equation}
I=\frac{k}{4\pi}\intop\langle AdA+\frac{2}{3}A^{3}\rangle\,,\label{eq:CSaction}
\end{equation}
which by virtue of \eqref{eq:bracketCov} and \eqref{eq:A}, reduces
to
\begin{equation}
I=\frac{k}{4\pi}\intop\left(2e^{a}R_{a}+i\bar{\psi}_{I}\nabla\psi_{I}-2BdC\right)\,,\label{eq:ActionHIPT}
\end{equation}
up to a boundary term. 

The level and the Newton constant are related as $k=\frac{1}{4G}$,
while $R^{a}=d\omega^{a}+\frac{1}{2}\epsilon^{abc}\omega_{b}\omega_{c}$
stands for the dualized curvature two-form. The covariant derivative
acting on spinors reads
\begin{equation}
\nabla\psi_{I}=d\psi_{I}+\frac{1}{2}\omega^{a}\Gamma_{a}\psi_{I}+B\epsilon^{IJ}\psi_{J}\,.
\end{equation}
The field equations then imply that the field strength
\begin{equation}
F=\tilde{T}^{a}P_{a}+R^{a}J_{a}+\nabla\psi_{I}^{\alpha}Q_{\alpha}^{I}+dB\,T+\tilde{F}_{C}\,Z
\end{equation}
vanishes, where $\tilde{T}^{a}$ and $\tilde{F}_{C}$ stand for the
supercovariant torsion and the supercovariant $U(1)$ curvature along
$Z$, respectively. They are given by
\begin{eqnarray}
\tilde{T}^{a} & = & T^{a}-\frac{1}{4}i\bar{\psi}_{I}\Gamma^{a}\psi_{I}\,,\\
\tilde{F}_{C} & = & dC+\frac{1}{2}i\epsilon^{IJ}\bar{\psi}_{I}\psi_{J}\,,
\end{eqnarray}
with $T^{a}=de^{a}+\epsilon^{abc}\omega_{b}e_{c}$. 

By construction, the action \eqref{eq:ActionHIPT} is invariant under
local supersymmetries that correspond to gauge transformations, $\delta A=d\lambda+\left[A,\lambda\right]$,
spanned by a Lie-algebra-valued fermionic parameter $\lambda=\epsilon_{I}^{\alpha}Q_{\alpha}^{I}$.
The nontrivial supersymmetry transformations of the fields are then
given by 
\begin{equation}
\delta e^{a}=\frac{1}{2}i\bar{\epsilon}_{I}\Gamma^{a}\psi_{I}\quad,\quad\delta\psi_{I}=\nabla\epsilon_{I}\quad,\quad\delta C=-i\epsilon^{IJ}\bar{\epsilon}_{I}\psi_{J}\,,\label{eq:localsusy}
\end{equation}
and therefore, along the lines of \cite{Banados:1996hi}, the algebra
of the local supersymmetries spanned in \eqref{eq:localsusy} can
be seen to close off-shell according to super-Poincaré with $\mathcal{N}=(2,0)$
in \eqref{eq:Cov-Algebra}, without the need of introducing auxiliary
fields.

In the next section we perform an exhaustive analysis of the asymptotic
structure of the theory. 

\section{Asymptotic structure: supersymmetric extension of BMS$_{3}$ with
$\mathcal{N}=(2,0)$\label{Asympt}}

The analysis of the asymptotic structure of $\mathcal{N}=1$ Poincaré
supergravity was carried out in \cite{Barnich:2014cwa}. Following
the lines of \cite{Henneaux:2013dra} and \cite{Bunster:2014mua},
it was extended in \cite{Fuentealba:2015wza} so as to incorporate
a generic choice of Lagrange multipliers at infinity (for the case
of fermionic fields of spin $s=\frac{3}{2}$). 

Here we extend these results to the case of Poincaré supergravity
with $\mathcal{N}=(2,0)$. It is then useful to change the basis of
the extended super-Poincaré algebra in \eqref{eq:Cov-Algebra} according
to
\begin{gather}
L_{-1}=-\sqrt{2}J_{0}\quad,\quad L_{1}=\sqrt{2}J_{1}\quad,\quad L_{0}=J_{2}\,,\nonumber \\
M_{-1}=-\sqrt{2}P_{0}\quad,\quad M_{1}=\sqrt{2}P_{1}\quad,\quad M_{0}=P_{2}\,,\label{eq:ChangeBasis}\\
G_{-\frac{1}{2}}^{I}=\sqrt{2}Q_{+}^{I}\quad,\quad G_{\frac{1}{2}}^{I}=\sqrt{2}Q_{-}^{I}\,,\nonumber 
\end{gather}
so that the superalgebra now reads
\begin{gather}
[L_{m},L_{n}]=\left(m-n\right)L_{m+n}\quad,\quad[L_{m},M_{n}]=\left(m-n\right)M_{m+n}\,,\nonumber \\{}
[L_{m},G_{p}^{I}]=\left(\frac{m}{2}-p\right)G_{m+p}^{I}\quad,\quad[G_{p}^{I},T]=\epsilon^{IJ}G_{p}^{J}\,,\label{eq:algSL2}\\
\{G_{p}^{I},G_{q}^{J}\}=\delta^{IJ}M_{p+q}-2\left(p-q\right)\epsilon^{IJ}Z\,,\nonumber 
\end{gather}
where $m,n=\pm1,0$, and $p,q=\pm\frac{1}{2}$. Thus, the nonvanishing
components of the invariant bilinear form in \eqref{eq:bracketCov}
reduce to
\begin{gather}
\langle L_{1},M_{-1}\rangle=\langle L_{-1},M_{1}\rangle=-2\quad,\quad\langle L_{0},M_{0}\rangle=1\,,\nonumber \\
\langle G_{-\frac{1}{2}}^{I},G_{\frac{1}{2}}^{J}\rangle=2\delta^{IJ}\quad,\quad\langle T,Z\rangle=-1\,.\label{eq:BracketSL2}
\end{gather}

In order to propose a suitable set of asymptotic conditions, it is
useful to follow some general criteria as the ones spelled out in
\cite{Henneaux:1985tv}, \cite{Henneaux:2006hk}, \cite{Henneaux:2009pw},
\cite{Henneaux:2010fy}, \cite{Perez:2015jxn}. Once adapted to the
theory under discussion, they are: 
\begin{enumerate}
\item[$\left(i\right)$]  The asymptotic symmetries of the set must include BMS$_{3}$ as
well as the two fermionic ones. 
\item[$\left(ii\right)$]  The fall-off of the fields has to be relaxed enough so as to incorporate
the bosonic solutions of interest. 
\item[$\left(iii\right)$]  The decay must simultaneously be sufficiently fast in order to ensure
finiteness of the variation of the global charges.
\item[$\left(iv\right)$]  The boundary conditions have to guarantee that the variation of
the charges fulfills suitable functional integrability conditions.
\end{enumerate}
Taking into account these four requirements, the asymptotic behaviour
is proposed to be of the form
\begin{equation}
A=h^{-1}ah+h^{-1}dh\,,\label{eq:Acontutti}
\end{equation}
where as in \cite{Coussaert:1995zp} the group element $h$ entirely
captures the dependence on radial coordinate $r$, so that the auxiliary
connection $a$ depends only on the remaining ones $u,\phi$.

In concrete, we choose $h=e^{\frac{r}{2}M_{-1}}$, and the spacelike
component of $a$ to be given by
\begin{equation}
a_{\phi}=L_{1}-\frac{\pi}{k}\left[\left(\mathcal{P}-\frac{4\pi}{k}\mathcal{Z}^{2}\right)L_{-1}+\left(\mathcal{J}+\frac{2\pi}{k}\mathcal{T}\mathcal{Z}\right)M_{-1}+\psi G_{-\frac{1}{2}}^{1}+\mathcal{S}G_{-\frac{1}{2}}^{2}+2\mathcal{Z}T+2\mathcal{T}Z\right]\,,\label{eq:aphi}
\end{equation}
so that deviations with respect to the background configuration, which
we assume to be given by the null orbifold \cite{Horowitz:1990ap},
are described by arbitrary functions of $u,\phi$ that go along the
highest weight generators. According to \cite{Henneaux:2013dra},
\cite{Bunster:2014mua}, the bosonic functions $\mathcal{P}$, $\mathcal{J}$,
$\mathcal{Z}$, $\mathcal{T}$, and the fermionic ones $\psi$, $\mathcal{S}$,
correspond to the dynamical fields. 

Moreover, as required by criterion $\left(ii\right)$, in order to
accommodate the widest possible class of bosonic solutions$,$ the
Lagrange multipliers associated to the dynamical fields have to be
explicitly incorporated in the asymptotic behaviour. They turn out
to be defined through the lowest weight components of
\begin{equation}
a_{u}=\Lambda[\mu_{\mathcal{J}},\mu_{\mathcal{P}},\mu_{\mathcal{\psi}},\mu_{\mathcal{S}},\mu_{\mathcal{T}},\mu_{\mathcal{Z}}]\,,\label{eq:au}
\end{equation}
 given by
\begin{eqnarray}
\Lambda & = & \mu_{\mathcal{J}}L_{1}+\mu_{\mathcal{P}}M_{1}+\mu_{\psi}G_{\frac{1}{2}}^{1}+\mu_{S}G_{\frac{1}{2}}^{2}+\left(\mu_{\mathcal{T}}-\frac{2\pi}{k}\mu_{\mathcal{J}}\mathcal{Z}\right)T+\left(\mu_{\mathcal{Z}}-\frac{2\pi}{k}\mu_{\mathcal{J}}\mathcal{T}+\frac{8\pi}{k}\mu_{\mathcal{P}}\mathcal{Z}\right)Z\nonumber \\
 &  & -\mu_{\mathcal{J}}\text{\ensuremath{'}}L_{0}-\mu_{\mathcal{P}}\text{\ensuremath{'}}M_{0}+\left[\frac{1}{2}\mu_{\mathcal{J}}\text{\ensuremath{''}}-\frac{\pi}{k}\mu_{\mathcal{J}}\left(\mathcal{P}-\frac{4\pi}{k}\mathcal{Z}^{2}\right)\right]L_{-1}\nonumber \\
 &  & +\left[\frac{1}{2}\mu_{\mathcal{P}}\text{\ensuremath{''}}-\frac{\pi}{k}\mu_{\mathcal{J}}\left(\mathcal{J}+\frac{2\pi}{k}\mathcal{T}\mathcal{Z}\right)-\frac{\pi}{k}\mu_{\mathcal{P}}\left(\mathcal{P}-\frac{4\pi}{k}\mathcal{Z}^{2}\right)+\frac{\pi}{2k}i\mathcal{\psi}\mu_{\mathcal{\psi}}+\frac{\pi}{2k}i\mathcal{S}\mu_{\mathcal{S}}\right]M_{-1}\nonumber \\
 &  & -\left(\mu_{\mathcal{\psi}}\text{\ensuremath{'}}+\frac{\pi}{k}\mu_{\mathcal{J}}\psi-\frac{2\pi}{k}\mathcal{Z}\mu_{\mathcal{S}}\right)G_{-\frac{1}{2}}^{1}-\left(\mu_{\mathcal{S}}\text{\ensuremath{'}}+\frac{\pi}{k}\mu_{\mathcal{J}}\mathcal{S}+\frac{2\pi}{k}\mathcal{Z}\mu_{\psi}\right)G_{-\frac{1}{2}}^{2}\,,\label{eq:Lambda}
\end{eqnarray}
where prime denotes $\partial_{\phi}$. 

The bosonic Lagrange multipliers $\mu_{\mathcal{J}}$, $\mu_{\mathcal{P}}$,
$\mu_{\mathcal{T}}$, $\mu_{\mathcal{Z}}$ as well as the fermionic
ones $\mu_{\psi}$, $\mu_{S}$, can be assumed to be given by arbitrary
independent functions of $u,\phi$, that are held fixed at the boundary
without variation.

The asymptotic symmetries are then described by the subset of gauge
transformations $\delta a=d\lambda+\left[a,\lambda\right]$ that preserve
the asymptotic form of the auxiliary connection in \eqref{eq:aphi},
\eqref{eq:au}. 

The spacelike component of $a$ in \eqref{eq:aphi} is maintained
for Lie-algebra-valued parameters of the form
\begin{equation}
\lambda=\Lambda[\epsilon_{\mathcal{J}},\epsilon_{\mathcal{P}},\epsilon_{\psi},\epsilon_{S},\epsilon_{\mathcal{T}},\epsilon_{\mathcal{Z}}]\,,\label{eq:Lambachico}
\end{equation}
provided that the transformation law of the dynamical fields is given
by
\begin{eqnarray}
\delta\mathcal{P} & = & 2\mathcal{P}\epsilon_{\mathcal{J}}\text{\ensuremath{'}}+\mathcal{P}\text{\ensuremath{'}}\epsilon_{\mathcal{J}}-\frac{k}{2\pi}\epsilon_{\mathcal{J}}\text{\ensuremath{'''}}-4\mathcal{Z}\epsilon_{\mathcal{T}}\text{\ensuremath{'}}\,,\nonumber \\
\delta\mathcal{J} & = & 2\mathcal{J}\epsilon_{\mathcal{J}}\text{\ensuremath{'}}+\mathcal{J}\text{\ensuremath{'}}\epsilon_{\mathcal{J}}+2\mathcal{P}\epsilon_{\mathcal{P}}\text{\ensuremath{'}}+\mathcal{P}\text{\ensuremath{'}}\epsilon_{\mathcal{P}}-\frac{k}{2\pi}\epsilon_{\mathcal{P}}\text{\ensuremath{'''}}\nonumber \\
 &  & +\mathcal{Z}\epsilon_{\mathcal{Z}}\text{\ensuremath{'}}+\mathcal{T}\epsilon_{\mathcal{T}}\text{\ensuremath{'}}-\frac{3}{2}i\psi\epsilon_{\mathcal{\psi}}\text{\ensuremath{'}}-\frac{1}{2}i\psi\text{\ensuremath{'}}\epsilon_{\mathcal{\psi}}-\frac{3}{2}i\mathcal{S}\epsilon_{\mathcal{S}}\text{\ensuremath{'}}-\frac{1}{2}i\mathcal{S}\text{\ensuremath{'}}\epsilon_{\mathcal{S}}\nonumber \\
\delta\psi & = & \frac{3}{2}\psi\epsilon_{\mathcal{J}}\text{\ensuremath{'}}+\psi\text{\ensuremath{'}}\epsilon_{\mathcal{J}}-\mathcal{S}\epsilon_{\mathcal{T}}-\mathcal{P}\epsilon_{\psi}+\frac{k}{\pi}\epsilon_{\psi}\text{\ensuremath{''}}-2\mathcal{Z}\text{\ensuremath{'}}\epsilon_{\mathcal{S}}-4\mathcal{Z}\epsilon_{\mathcal{S}}\text{\ensuremath{'}}\,,\label{eq:delta}\\
\delta\mathcal{S} & = & \frac{3}{2}\mathcal{S}\epsilon_{\mathcal{J}}\text{\ensuremath{'}}+\mathcal{S}\text{\ensuremath{'}}\epsilon_{\mathcal{J}}+\psi\epsilon_{\mathcal{T}}-\mathcal{P}\epsilon_{\mathcal{S}}+\frac{k}{\pi}\epsilon_{\mathcal{S}}\text{\ensuremath{''}}+2\mathcal{Z}\text{\ensuremath{'}}\epsilon_{\psi}+4\mathcal{Z}\epsilon_{\psi}\text{\ensuremath{'}}\,,\nonumber \\
\delta\mathcal{T} & = & -i\mathcal{S}\epsilon_{\psi}+i\mathcal{\psi}\epsilon_{\mathcal{S}}-\frac{k}{2\pi}\epsilon_{\mathcal{Z}}\text{\ensuremath{'}}+\epsilon_{\mathcal{J}}\text{\ensuremath{'}}\mathcal{T}+\epsilon_{\mathcal{J}}\mathcal{T}\text{\ensuremath{'}}-4\epsilon_{\mathcal{P}}\text{\ensuremath{'}}\mathcal{Z}-4\epsilon_{\mathcal{P}}\mathcal{Z}\text{\ensuremath{'}}\,,\nonumber \\
\delta\mathcal{Z} & = & -\frac{k}{2\pi}\epsilon_{\mathcal{T}}\text{\ensuremath{'}}+\epsilon_{\mathcal{J}}\text{\ensuremath{'}}\mathcal{Z}+\epsilon_{\mathcal{J}}\mathcal{Z}\text{\ensuremath{'}}\,.\nonumber 
\end{eqnarray}
Note that $\Lambda$ in \eqref{eq:Lambachico} is precisely the same
as in \eqref{eq:Lambda}, but now depends on arbitrary bosonic and
fermionic functions of $u,\phi$, given by $\epsilon_{\mathcal{J}}$,
$\epsilon_{\mathcal{P}}$, $\epsilon_{\mathcal{T}}$, $\epsilon_{\mathcal{Z}}$,
and $\epsilon_{\psi}$, $\epsilon_{S}$, respectively.

Preserving the form of $a_{u}$ then implies that the field equations
have to hold at the asymptotic region, and also provides additional
suitable conditions for the parameters that span the asymptotic symmetries.
In the reduce phase space, the field equations then read
\begin{eqnarray}
\dot{\mathcal{P}} & = & 2\mathcal{P}\mu_{\mathcal{J}}\text{\ensuremath{'}}+\mathcal{P}\text{\ensuremath{'}}\mu_{\mathcal{J}}-\frac{k}{2\pi}\mu_{\mathcal{J}}\text{\ensuremath{'''}}-4\mathcal{Z}\mu_{\mathcal{T}}\text{\ensuremath{'}}\,,\nonumber \\
\dot{\mathcal{J}} & = & 2\mathcal{J}\mu_{\mathcal{J}}\text{\ensuremath{'}}+\mathcal{J}\text{\ensuremath{'}}\mu_{\mathcal{J}}+2\mathcal{P}\mu_{\mathcal{P}}\text{\ensuremath{'}}+\mathcal{P}\text{\ensuremath{'}}\mu_{\mathcal{P}}-\frac{k}{2\pi}\mu_{\mathcal{P}}\text{\ensuremath{'''}}\nonumber \\
 &  & +\mathcal{Z}\mu_{\mathcal{Z}}\text{\ensuremath{'}}+\mathcal{T}\mu_{\mathcal{T}}\text{\ensuremath{'}}-\frac{3}{2}i\psi\mu_{\mathcal{\psi}}\text{\ensuremath{'}}-\frac{1}{2}i\psi\text{\ensuremath{'}}\mu_{\mathcal{\psi}}-\frac{3}{2}i\mathcal{S}\mu_{\mathcal{S}}\text{\ensuremath{'}}-\frac{1}{2}i\mathcal{S}\text{\ensuremath{'}}\mu_{\mathcal{S}}\nonumber \\
\dot{\psi} & = & \frac{3}{2}\psi\mu_{\mathcal{J}}\text{\ensuremath{'}}+\psi\text{\ensuremath{'}}\mu_{\mathcal{J}}-\mathcal{S}\mu_{\mathcal{T}}-\mathcal{P}\mu_{\psi}+\frac{k}{\pi}\mu_{\psi}\text{\ensuremath{''}}-2\mathcal{Z}\text{\ensuremath{'}}\mu_{\mathcal{S}}-4\mathcal{Z}\mu_{\mathcal{S}}\text{\ensuremath{'}}\,,\label{eq:EoM}\\
\dot{\mathcal{S}} & = & \frac{3}{2}\mathcal{S}\mu_{\mathcal{J}}\text{\ensuremath{'}}+\mathcal{S}\text{\ensuremath{'}}\mu_{\mathcal{J}}+\psi\mu_{\mathcal{T}}-\mathcal{P}\mu_{\mathcal{S}}+\frac{k}{\pi}\mu_{\mathcal{S}}\text{\ensuremath{''}}+2\mathcal{Z}\text{\ensuremath{'}}\mu_{\psi}+4\mathcal{Z}\mu_{\psi}\text{\ensuremath{'}}\,,\nonumber \\
\dot{\mathcal{T}} & = & -i\mathcal{S}\mu_{\psi}+i\mathcal{\psi}\mu_{\mathcal{S}}-\frac{k}{2\pi}\mu_{\mathcal{Z}}\text{\ensuremath{'}}+\mu_{\mathcal{J}}\text{\ensuremath{'}}\mathcal{T}+\mu_{\mathcal{J}}\mathcal{T}\text{\ensuremath{'}}-4\mu_{\mathcal{P}}\text{\ensuremath{'}}\mathcal{Z}-4\mu_{\mathcal{P}}\mathcal{Z}\text{\ensuremath{'}}\,,\nonumber \\
\dot{\mathcal{Z}} & = & -\frac{k}{2\pi}\mu_{\mathcal{T}}\text{\ensuremath{'}}+\mu_{\mathcal{J}}\text{\ensuremath{'}}\mathcal{Z}+\mu_{\mathcal{J}}\mathcal{Z}\text{\ensuremath{'}}\,,\nonumber 
\end{eqnarray}
which can be readily obtained from \eqref{eq:delta} by taking into
account that time evolution is generated by gauge transformations
whose parameters correspond to the Lagrange multipliers. The conditions
for the parameters are explicitly given by
\begin{eqnarray}
\dot{\epsilon}_{\mathcal{J}} & = & \mu_{\mathcal{J}}\epsilon_{\mathcal{J}}\text{\ensuremath{'}}-\mu_{\mathcal{J}}\text{\ensuremath{'}}\epsilon_{\mathcal{J}}\,,\nonumber \\
\dot{\epsilon}_{\mathcal{P}} & = & \mu_{\mathcal{P}}\epsilon_{\mathcal{J}}\text{\ensuremath{'}}+\mu_{\mathcal{J}}\epsilon_{\mathcal{P}}\text{\ensuremath{'}}-\mu_{\mathcal{J}}\text{\ensuremath{'}}\epsilon_{\mathcal{P}}-\mu_{\mathcal{P}}\text{\ensuremath{'}}\epsilon_{\mathcal{J}}-i\mu_{\mathcal{S}}\epsilon_{\mathcal{S}}-i\mu_{\psi}\epsilon_{\psi}\,,\nonumber \\
\dot{\epsilon}_{\psi} & = & \mu_{\mathcal{J}}\epsilon_{\psi}\text{\ensuremath{'}}+\frac{1}{2}\mu_{\psi}\epsilon_{\mathcal{J}}\text{\ensuremath{'}}-\frac{1}{2}\mu_{\mathcal{J}}\text{\ensuremath{'}}\epsilon_{\psi}-\mu_{\psi}\text{\ensuremath{'}}\epsilon_{\mathcal{J}}+\mu_{\mathcal{S}}\epsilon_{\mathcal{T}}-\mu_{\mathcal{T}}\epsilon_{\mathcal{S}}\,,\nonumber \\
\dot{\epsilon}_{\mathcal{S}} & = & \mu_{\mathcal{J}}\epsilon_{\mathcal{S}}\text{\ensuremath{'}}+\frac{1}{2}\mu_{\mathcal{S}}\epsilon_{\mathcal{J}}\text{\ensuremath{'}}-\frac{1}{2}\mu_{\mathcal{J}}\text{\ensuremath{'}}\epsilon_{\mathcal{S}}-\mu_{\mathcal{S}}\text{\ensuremath{'}}\epsilon_{\mathcal{J}}+\mu_{\mathcal{T}}\epsilon_{\psi}-\mu_{\psi}\epsilon_{\mathcal{T}}\,,\label{eq:Chiralities}\\
\dot{\epsilon}_{\mathcal{Z}} & = & \mu_{\mathcal{J}}\epsilon_{\mathcal{Z}}\text{\ensuremath{'}}-\mu_{\mathcal{Z}}\text{\ensuremath{'}}\epsilon_{\mathcal{J}}-4\mu_{\mathcal{P}}\epsilon_{\mathcal{T}}\text{\ensuremath{'}}+4\mu_{\mathcal{T}}\text{\ensuremath{'}}\epsilon_{\mathcal{P}}-2i\mu_{\psi}\epsilon_{\mathcal{S}}\text{\ensuremath{'}}+2i\mu_{\mathcal{S}}\epsilon_{\psi}\text{\ensuremath{'}}-2i\mu_{\mathcal{S}}\text{\ensuremath{'}}\epsilon_{\psi}+2i\mu_{\psi}\text{\ensuremath{'}}\epsilon_{\mathcal{S}}\,,\nonumber \\
\dot{\epsilon}_{\mathcal{T}} & = & \mu_{\mathcal{J}}\epsilon_{\mathcal{T}}\text{\ensuremath{'}}-\mu_{\mathcal{T}}\text{\ensuremath{'}}\epsilon_{\mathcal{J}}\,,\nonumber 
\end{eqnarray}
and they can be seen to ensure that the variation of the canonical
generators is conserved. This is discussed next.

\subsection{Canonical generators and their algebra}

In the canonical approach \cite{Regge:1974zd}, the variation of the
surface integrals that define the global charges is given by
\begin{equation}
\delta Q[\lambda]=-\frac{k}{2\pi}\int\langle\lambda\delta a_{\phi}\rangle d\phi\,,
\end{equation}
which by virtue of \eqref{eq:BracketSL2}, \eqref{eq:aphi}, \eqref{eq:Lambachico},
readily integrates as 
\begin{equation}
Q[\epsilon_{\mathcal{J}},\epsilon_{\mathcal{P}},\epsilon_{\psi},\epsilon_{S},\epsilon_{\mathcal{T}},\epsilon_{\mathcal{Z}}]=-\int\left(\epsilon_{\mathcal{J}}\mathcal{J}+\epsilon_{\mathcal{P}}\mathcal{P}+i\epsilon_{\mathcal{\psi}}\mathcal{\psi}+i\epsilon_{\mathcal{S}}\mathcal{S}+\epsilon_{\mathcal{T}}\mathcal{T}+\epsilon_{\mathcal{Z}}\mathcal{Z}\right)d\phi\,.\label{eq:Q}
\end{equation}
The asymptotic symmetry algebra can then be obtained from the direct
evaluation of the Poisson brackets of the global charges in \eqref{eq:Q}.
As a shortcut, if one takes into account the variation of the dynamical
fields in \eqref{eq:delta}, the asymptotic symmetry algebra can be
directly read from $\delta_{\lambda_{2}}Q[\lambda_{1}\}=\{Q\left[\lambda_{1}\right],Q\left[\lambda_{2}\right]$\},
so that once expanding in Fourier modes according to $X_{n}=\int Xe^{-in\phi}d\phi$,
it is given by
\begin{eqnarray}
i\{\mathcal{J}_{m},\mathcal{J}_{n}\} & = & \left(m-n\right)\mathcal{J}_{m+n}\,,\nonumber \\
i\{\mathcal{J}_{m},\mathcal{P}_{n}\} & = & \left(m-n\right)\mathcal{P}_{m+n}+km^{3}\delta_{m+n,0}\,,\nonumber \\
i\{\mathcal{J}_{m},\mathcal{G}_{p}^{I}\} & = & \left(\frac{m}{2}-p\right)\mathcal{G}_{m+p}^{I}\,,\nonumber \\
i\{\mathcal{\mathcal{J}}_{m},\mathcal{Z}_{n}\} & = & -n\mathcal{Z}_{m+n}\,,\nonumber \\
i\{\mathcal{J}_{m},\mathcal{T}_{n}\} & = & -n\mathcal{T}_{m+n}\,,\label{eq:alg3}\\
i\{\mathcal{P}_{m},\mathcal{T}_{n}\} & = & 4n\mathcal{Z}_{m+n}\,,\nonumber \\
i\{\mathcal{T}_{m},\mathcal{Z}_{n}\} & = & -km\delta_{m+n,0}\,,\nonumber \\
i\{\mathcal{G}_{p}^{I},\mathcal{T}_{m}\} & = & i\epsilon^{IJ}\mathcal{G}_{m+p}^{J}\,,\nonumber \\
i\{\mathcal{G}_{p}^{I},\mathcal{G}_{q}^{J}\} & = & \delta^{IJ}\left(\mathcal{P}_{p+q}+2kp^{2}\delta_{p+q,0}\right)+2i\epsilon^{IJ}\left(p-q\right)\mathcal{Z}_{p+q}\,,\nonumber 
\end{eqnarray}
where $\mathcal{G}^{1}=\psi$ and $\mathcal{G}^{2}=\mathcal{S}$. 

Here, $m$, $n$ stand for integers, and $p$, $q$ are given by (half-)integers
when fermions fulfill (anti)periodic boundary conditions, while the
reality conditions of the modes are given by $\ensuremath{(\mathcal{J}_{m})^{*}=\mathcal{J}_{-m}}$,
$\ensuremath{(\mathcal{P}_{m})^{*}=\mathcal{P}_{-m}}$, $(\mathcal{T}_{m})^{*}=\mathcal{T}_{-m}$,
$\ensuremath{(\mathcal{Z}_{m})^{*}=\mathcal{Z}_{-m}}$, $\ensuremath{(\mathcal{G}_{m}^{I})^{*}=\mathcal{G}_{-m}^{I}}$.

The asymptotic symmetry algebra \eqref{eq:alg3} then manifestly contains
BMS$_{3}$ with the central extension found in \cite{Barnich:2006av}
as the subalgebra spanned by $\mathcal{J}_{m}$ and $\mathcal{P}_{m}$.
The remaining part of the bosonic subalgebra\footnote{As an interesting remark, it is worth pointing that the bosonic subalgebra
of \eqref{eq:alg3} coincides with the one obtained in \cite{Detournay:2016sfv}
for pure gravity with a nonstandard set of boundary conditions, which
also corresponds to the vanishing cosmological constant limit of the
asymptotic symmetry algebra found in \cite{Troessaert:2013fma}.} consists on two commuting independent affine algebras generated by
the spin-one currents $J_{m}^{\pm}=\frac{1}{\sqrt{2}}\left(\mathcal{Z}_{m}\pm\mathcal{T}_{m}\right)$,
whose level is determined by $k$.

The super-Poincaré algebra in \eqref{eq:algSL2} tuns out to be a
subalgebra of \eqref{eq:alg3} in the antiperiodic case, being spanned
by $\left\{ \mathcal{J}_{m},\mathcal{P}_{n},\mathcal{G}_{p}^{\left(i\right)},\text{\ensuremath{\mathcal{Z}}}_{0},\text{\ensuremath{\mathcal{T}}}_{0}\right\} $
provided the labels are restricted according to $m,n=\pm1,0$ and
$p,q=\pm\frac{1}{2}$, and the supertranslation generator $\mathcal{P}_{0}$
is shifted as $\mathcal{P}_{0}\rightarrow\mathcal{P}_{0}+\frac{k}{2}$.
The explicit matching is recovered provided that $L_{m}=\mathcal{J}_{m}$,
$M_{m}=\mathcal{P}_{m}$, $G_{p}^{\left(i\right)}=\mathcal{G}_{p}^{\left(i\right)}$,
$T=-\mathcal{T}_{0}$ and $Z=-\mathcal{Z}_{0}$.

\subsubsection{Asymptotic symmetry algebra from a truncation of the homogeneous
contraction of the superconformal algebra with $\mathcal{N}=(2,2)$}

It is simple to verify that the supersymmetric extension of the BMS$_{3}$
algebra with $\mathcal{N}=(2,0)$ in \eqref{eq:alg3}, can be recovered
from a homogeneous (democratic) or ultra-relativistic Inönü-Wigner
contraction of the superconformal algebra with $\mathcal{N}=(2,2)$,
provided that the fermionic generators of the right copy are truncated
(see e.g., \cite{Banerjee:2016nio}). Indeed, this is so regardless
the truncation is performed before or after the contraction process.
This can be seen as follows.

The superconformal algebra with $\mathcal{N}=(2,2)$ is given by two
independent (left and right) copies of the super-Virasoro algebra
with $\mathcal{N}=2$, which reads 
\begin{eqnarray}
i\{\mathcal{L}_{m},\mathcal{L}_{n}\} & = & \left(m-n\right)\mathcal{L}_{m+n}+\frac{\kappa}{2}m^{3}\delta_{m+n,0}\,,\\
i\{\mathcal{L}_{m},\mathcal{Q}_{p}^{I}\} & = & \left(\frac{m}{2}-p\right)\mathcal{Q}_{m+p}^{I}\,,\\
i\{\mathcal{L}_{m},\mathcal{R}_{n}\} & = & -n\mathcal{R}_{m+n}\,,\\
i\{\mathcal{R}_{m},\mathcal{R}_{n}\} & = & 2\kappa m\delta_{m+n,0}\,,\\
i\{\mathcal{Q}_{p}^{I},\mathcal{R}_{m}\} & = & -i\epsilon^{IJ}\mathcal{Q}_{m+p}^{J}\,,\\
i\{\mathcal{Q}_{p}^{I},\mathcal{Q}_{q}^{J}\} & = & \delta^{IJ}\left(\mathcal{L}_{p+q}+\kappa p^{2}\delta_{p+q,0}\right)+\frac{1}{2}i\epsilon^{IJ}(p-q)\mathcal{R}_{p+q}\,.
\end{eqnarray}
It is then useful to change the basis according to
\begin{gather}
\mathcal{J}_{m}=\mathcal{L}_{m}^{+}-\mathcal{L}_{-m}^{-}\qquad,\qquad\mathcal{P}_{m}=\frac{1}{\ell}\left(\mathcal{L}_{m}^{+}+\mathcal{L}_{-m}^{-}\right)\,,\\
\mathcal{T}_{m}=-\left(\mathcal{R}_{m}^{+}-\mathcal{R}_{-m}^{-}\right)\qquad,\qquad\mathcal{Z}_{m}=\frac{1}{4\ell}\left(\mathcal{R}_{m}^{+}+\mathcal{R}_{-m}^{-}\right)\,,\\
\mathcal{G}_{p}^{+I}=\sqrt{\frac{2}{\ell}}\mathcal{Q}_{r}^{+I}\qquad,\qquad\mathcal{G}_{p}^{-I}=\sqrt{\frac{2}{\ell}}\mathcal{Q}_{-r}^{-I}\,,
\end{gather}
so that the contraction process is performed through rescaling the
level as $\kappa=k\ell$, and then taking the limit $\ell\rightarrow\infty$.
It is then clear that if one truncates the fermionic generators of
the right copy, the super-BMS$_{3}$ algebra with $\mathcal{N}=(2,0)$
in \eqref{eq:alg3} is recovered with $\mathcal{G}_{p}^{I}=\mathcal{G}_{p}^{+I}$.

As a final remark of this section, it is also worth pointing out that
there is a different, so-called inhomogeneous, or ``despotic'', Inönü-Wigner
contraction of the superconformal algebra with $\mathcal{N}=(2,2)$,
which leads to an inequivalent supersymmetric extension of the super-BMS$_{3}$
algebra with $\mathcal{N}=(2,0)$. This alternative possibility has
been simultaneously analyzed in the context of the asymptotic structure
of a different theory in \cite{Basu:2017aqn}. 

\section{Nonlinear energy bounds from super-BMS$_{3}$ with $\mathcal{N}=(2,0)$
\label{Bounds}}

Supersymmetric bounds for different definitions of the global charges
in the context of the supergravity theory under consideration, have
been previously discussed in \cite{Howe:1995zm}, \cite{Edelstein:1995md}.
In this section we show that the Poisson brackets of the fermionic
generators of the asymptotic supersymmetries in \eqref{eq:alg3},
in spite of being linear in the bosonic generators, yield an infinite
number of nonlinear bounds for the energy. Only a finite number of
them are able to saturate, precisely corresponding to the same number
of unbroken supersymmetries. In order to carry out this task, as explained
in \cite{Bunster:2014mua}, generic bosonic configurations can be
brought to the \textquotedblleft rest frame\textquotedblright{} by
acting on them with a suitable combination of the asymptotic symmetries.
Therefore, it is enough to focus in the case of bosonic configurations
endowed with zero mode charges, given by $\mathcal{P}_{0}=2\pi\mathcal{P}$
and $\mathcal{Z}_{0}=2\pi\mathcal{Z}$, regardless the value of the
remaining ones ($\mathcal{J}$ and $\mathcal{T}$). The supersymmetric
bounds we look for can then be obtained along the semi-classical reasoning
in \cite{Deser-Teitelboim}, \cite{Teitelboim_Susy}, \cite{Witten-positivity},
\cite{Abott-Deser}, \cite{Hull:1983ap}, \cite{Teitelboim:1984kf},
\cite{Coussaert-Henneaux}. In particular, we follow a similar strategy
as the one in \cite{Henneaux:2015tar}. The fermionic Poisson brackets
in \eqref{eq:alg3} are then promoted to anticommutators. In the case
of $p=-q=r$ they become

\begin{equation}
\left(2\pi\right)^{-1}\left(\hat{\mathcal{G}}_{r}^{I}\hat{\mathcal{G}}_{-r}^{J}+\hat{\mathcal{G}}_{-r}^{J}\hat{\mathcal{G}}_{r}^{I}\right)=\delta^{IJ}\left(\hat{\mathcal{P}}+\frac{k}{\pi}r^{2}\right)+4ir\epsilon^{IJ}\hat{\mathcal{Z}}=B_{r}^{IJ}\,,\label{eq:BrIJ}
\end{equation}
so that $(B_{r}^{IJ})^{\dagger}=B_{r}^{JI}$. 

Therefore, when $I=J$, the left hand side of \eqref{eq:BrIJ} is
a positive definite hermitian operator for any value of $r$ and $I$,
which implies that in the classical limit, the bosonic charges have
to fulfill the following bounds

\begin{equation}
B_{r}^{II}=\frac{k}{\pi}r^{2}+\mathcal{P}\geq0\,.\label{eq:BIIr}
\end{equation}
In the case of periodic boundary conditions for the fermionic fields,
the strongest bound in \eqref{eq:BIIr} corresponds to $r=0$, which
implies that the energy is nonnegative ($\mathcal{P}\geq0$). Analogously,
for antiperiodic boundary conditions, the strongest bound is given
by $r=\pm\frac{1}{2}$, so that the energy becomes bounded from below
according to $\mathcal{P}\geq-\frac{k}{4\pi}$.

Additional bounds also arise in the case of $I\neq J$. In order to
obtain them, it is useful to define the following complex fields
\begin{equation}
\hat{\mathcal{G}}_{r}^{\pm}:=\frac{1}{\sqrt{2}}\left(\hat{\mathcal{G}}_{r}^{1}\pm i\hat{\mathcal{G}}_{r}^{2}\right)\,,
\end{equation}
which fulfill $\hat{\mathcal{G}}_{r}^{\pm}\left(\hat{\mathcal{G}}_{r}^{\pm}\right)^{\dagger}+\left(\hat{\mathcal{G}}_{r}^{\pm}\right)^{\dagger}\hat{\mathcal{G}}_{r}^{\pm}\geq0$.
The latter bounds then imply that $B_{r}^{11}\geq\pm iB_{r}^{12}$,
which by virtue of \eqref{eq:BrIJ} yields
\begin{equation}
r^{2}\pm\frac{4\pi}{k}r\mathcal{Z}+\frac{\pi}{k}\mathcal{P}\geq0\,.\label{eq:bound2}
\end{equation}
It is convenient to factorize the bounds in \eqref{eq:bound2} according
to
\begin{equation}
\left(r\pm\lambda_{[+]}\right)\left(r\pm\lambda_{[-]}\right)\geq0\,,\label{eq:Bpmr}
\end{equation}
with
\begin{equation}
\lambda_{[\pm]}=-\frac{2\pi}{k}\mathcal{Z}\pm\sqrt{\frac{\pi}{k}\left(\frac{4\pi}{k}\mathcal{Z}^{2}-\mathcal{P}\right)}\,,\label{eq:lambdapm}
\end{equation}
where $\Lambda_{[\pm]}=i\lambda_{[\pm]}$ correspond to the eigenvalues
of the spacelike components of the $sl(2,R)\oplus u(1)$ connection
$\hat{\omega}=\omega+B$ that minimally couples to the fermionic fields
(see section eq. \eqref{eq:Lambdapm} below). 

It is then clear from \eqref{eq:lambdapm} that for periodic or antiperiodic
boundary conditions, in the case of $\mathcal{P}-\frac{4\pi}{k}\mathcal{Z}^{2}>0$,
the bounds in \eqref{eq:Bpmr} are automatically fulfilled and never
saturated. 

In the remaining possibility, $\mathcal{P}-\frac{4\pi}{k}\mathcal{Z}^{2}\leq0$,
the bounds in \eqref{eq:Bpmr} can be satisfied only provided that
$\lambda_{[+]}-\lambda_{[-]}\leq1$, which by virtue of \eqref{eq:lambdapm}
implies $\mathcal{P}\geq-\frac{k}{4\pi}+\frac{4\pi}{k}\mathcal{Z}^{2}$.
\newline

In sum, taking into account the infinite number of bounds in \eqref{eq:BIIr}
and \eqref{eq:Bpmr}, one deduces that the stronger ones imply that
energy has to be bounded from below as follows.
\begin{enumerate}
\item[] Antiperiodic boundary conditions: $\mathcal{P}\geq-\frac{k}{4\pi}+\frac{4\pi}{k}\mathcal{Z}^{2}$
\item[] Periodic boundary conditions: $\mathcal{P}\geq\begin{cases}
\begin{array}{cc}
0 & ,\\
-\frac{k}{4\pi}+\frac{4\pi}{k}\mathcal{Z}^{2} & ,
\end{array} & \begin{array}{c}
|\mathcal{Z}|<\frac{k}{4\pi}\\
|\mathcal{Z}|\geq\frac{k}{4\pi}
\end{array}\end{cases}$
\end{enumerate}
As a closing remark of this section, it is worth emphasizing that,
although the algebra of the asymptotic symmetries is a linear one,
the lower bounds for energy turn out to be quadratic in the electric-like
$\hat{u}(1)$ charge.

\section{Bosonic configurations and some of their properties \label{Configurations}}

Here we explore some properties of stationary spherically symmetric
bosonic solutions that fit within the asymptotic fall off described
in section \ref{Asympt}. The solutions are endowed just with zero-mode
bosonic charges, and hence they are described through dynamical gauge
fields given by
\begin{equation}
a_{\phi}=L_{1}-\frac{\pi}{k}\left[\left(\mathcal{P}-\frac{4\pi}{k}\mathcal{Z}^{2}\right)L_{-1}+\left(\mathcal{J}+\frac{2\pi}{k}\mathcal{T}\mathcal{Z}\right)M_{-1}+2\mathcal{Z}T+2\mathcal{T}Z\right]\,,\label{eq:aphi-bosonic}
\end{equation}
with $\mathcal{J}$, $\mathcal{P}$, $\mathcal{T}$, $\mathcal{Z}$,
constants. 

In the absence of fermionic charges, it is consistent to switch off
their corresponding Lagrange multipliers; i.e., the \textquotedblleft chemical
potentials\textquotedblright{} $\mu_{\psi}$, $\mu_{S}$, can be set
to vanish. For the sake of simplicity, the remaining bosonic ones,
given by $\mu_{\mathcal{J}}$, $\mu_{\mathcal{P}}$, $\mu_{\mathcal{T}}$,
$\mu_{\mathcal{Z}}$, are assumed to be constants and held fixed without
variation at the boundary. Therefore, the timelike component of the
gauge field in \eqref{eq:au} reduces to
\begin{align}
a_{u} & =\mu_{\mathcal{J}}L_{1}+\mu_{\mathcal{P}}M_{1}+\left(\mu_{\mathcal{T}}-\frac{2\pi}{k}\mu_{\mathcal{J}}\mathcal{Z}\right)T+\left(\mu_{\mathcal{Z}}-\frac{2\pi}{k}\mu_{\mathcal{J}}\mathcal{T}+\frac{8\pi}{k}\mu_{\mathcal{P}}\mathcal{Z}\right)Z\nonumber \\
 & -\left[\frac{\pi}{k}\mu_{\mathcal{J}}\left(\mathcal{J}+\frac{2\pi}{k}\mathcal{T}\mathcal{Z}\right)+\frac{\pi}{k}\mu_{\mathcal{P}}\left(\mathcal{P}-\frac{4\pi}{k}\mathcal{Z}^{2}\right)\right]M_{-1}-\frac{\pi}{k}\mu_{\mathcal{J}}\left(\mathcal{P}-\frac{4\pi}{k}\mathcal{Z}^{2}\right)L_{-1}\,.\label{eq:au-bosonic}
\end{align}

\subsection{Regularity conditions \label{Regularity}}

In the case of regular solutions, their smoothness can be established
through the fact that the holonomy of the gauge fields along a contractible
cycle $\mathcal{C}$ has to be trivial, i.e., 
\begin{equation}
H_{\mathcal{C}}=Pe^{\intop_{\mathcal{C}}a_{\mu}dx^{\mu}}=\Gamma^{\pm}\,,\label{eq:Hc}
\end{equation}
where the sign of $\Gamma^{\pm}$ corresponds to different choices
of spin structures (see, e.g, \cite{Henneaux:2015ywa}). Indeed, $\Gamma^{+}$
belongs to the center of the group, but this is not the case for $\Gamma^{-}$,
since for antiperiodic boundary condition it must anticommute with
the fermionic generators, i.e., $\{\Gamma^{-},G_{p}^{I}\}=0$.

In order to evaluate the regularity condition \eqref{eq:Hc}, it is
worth pointing out that neither the Poincaré algebra nor the super-Poincaré
algebra with $\mathcal{N}=(2,0)$ in \eqref{eq:Cov-Algebra} possess
a suitable standard matrix representation from which neither the invariant
bilinear form in \eqref{eq:bracketCov} nor the Casimir operator in
\eqref{eq:Casimir} can be obtained from the trace of a product of
two generators. Hence, regularity cannot be directly carried out through
``diagonalizing'' the holonomy in \eqref{eq:Hc}. Nonetheless, this
task might be carried out by virtue of a nonstandard matrix representation
\cite{Krishnan:2013wta} along the lines of \cite{Gary:2014ppa},
\cite{Riegler:2016hah}. Hereafter, regularity conditions are implemented
according to \cite{Matulich:2014hea}, which possesses the advantage
of being independent of the existence of a suitable matrix representation
for the entire gauge group. Once adapted to super-Poincaré with $\mathcal{N}=(2,0)$,
one proceeds as follows:
\begin{enumerate}
\item[$\left(i\right)$]  One begins finding a group element of the form $g=e^{\lambda_{n}M_{n}+\lambda Z}$
that allows gauging away the components of the gauge field along $M_{n}$
and $Z$ projected over the contractible cycle, so that the corresponding
components of the dreibein $e$ and the $U(1)$ field $C$ can be
consistently set to vanish. Consequently, the Lagrange multipliers
of electric type become generically fixed in terms of the ones of
magnetic type and the global charges.
\item[$\left(ii\right)$]  Regularity conditions can then be straightforwardly evaluated through
the diagonalization of the holonomy matrix along the contractible
cycle for the fundamental representation of the remaining $sl(2,R)\oplus u(1)$
connection $\hat{\omega}=\omega+B$.
\end{enumerate}
In order to continue with the analysis of the bosonic solutions under
consideration, it is necessary to identify the suitable contractible
cycles along which the regularity conditions have to be applied. Indeed,
these cycles might correspond either to the circles along Euclidean
time or the ones for the angular coordinate. These are the cases of
cosmological spacetimes or solitonic-like solutions, respectively.
This is discussed in what follows.

\subsection{Metric formalism}

The spacetime metric can be readily constructed from identifying the
dreibein from the components of the full gauge field $A$ in \eqref{eq:Acontutti}
along the generators $P_{a}$. The dreibein then reads
\begin{eqnarray}
e^{0} & = & \frac{\sqrt{2}\pi}{k}\left(\mathcal{J}+\frac{2\pi}{k}\mathcal{T}\mathcal{Z}\right)\left(d\phi+\mu_{\mathcal{J}}du\right)-\frac{1}{\sqrt{2}}dr+\frac{\sqrt{2}\pi}{k}\left(\mathcal{P}-\frac{4\pi}{k}\mathcal{Z}^{2}\right)\mu_{\mathcal{P}}du\,,\\
e^{1} & = & \sqrt{2}\mu_{\mathcal{P}}du\,,\\
e^{2} & = & r\left(d\phi+\mu_{\mathcal{J}}du\right)\,,
\end{eqnarray}
so that the line element is recovered in outgoing null coordinates,
as (see e.g., \cite{Gary:2014ppa}, \cite{Matulich:2014hea})
\begin{equation}
ds^{2}=-\frac{4\pi}{k}\left(\frac{\pi\mathcal{N}^{2}}{kr^{2}}-\mathcal{M}\right)\mu_{\mathcal{P}}^{2}du^{2}-2\mu_{\mathcal{P}}drdu+r^{2}\left[d\phi+\left(\mu_{\mathcal{J}}+\frac{2\pi\mu_{\mathcal{P}}\mathcal{N}}{kr^{2}}\right)du\right]^{2}\,,\label{eq:Line-Element}
\end{equation}
where the integration constants $\mathcal{M}$, $\mathcal{N}$ are
related to the global charges according to eqs. \eqref{eq:Energy},
\eqref{eq:AngMom}. The entire configuration is then determined by
the metric in \eqref{eq:Line-Element} together with the $U(1)$ fields
of electric and magnetic type, given by
\begin{eqnarray}
B & = & -\frac{2\pi}{k}\mathcal{Z}\left(d\phi+\mu_{\mathcal{J}}du\right)+\mu_{\mathcal{T}}du\,,\\
C & = & -\frac{2\pi}{k}\mathcal{T}\left(d\phi+\mu_{\mathcal{J}}du\right)+\left(\mu_{\mathcal{Z}}+\frac{8\pi}{k}\mu_{\mathcal{P}}\mathcal{Z}\right)du\,,
\end{eqnarray}
respectively.

It is worth pointing out that the spacetime metric does not acquire
a back reaction due to the presence of the $U(1)$ fields. Indeed,
as it can be seen from the action in \eqref{eq:ActionHIPT}, this
has to be so because, as they are described by a ``BF''-type of Lagrangian,
they do not couple to the metric and hence they cannot contribute
to the stress-energy tensor. Nonetheless, their presence does not
go unnoticed because they manifestly contribute to the energy and
the angular momentum of the configuration, given by
\begin{eqnarray}
\mathcal{P} & = & \mathcal{M}+\frac{4\pi}{k}\mathcal{Z}^{2}\,,\label{eq:Energy}\\
\mathcal{J} & = & \mathcal{N}-\frac{2\pi}{k}\mathcal{T}\mathcal{Z}\,,\label{eq:AngMom}
\end{eqnarray}
where $\mathcal{Z}$ and $\mathcal{T}$ stand for the $U(1)$ charges
of electric and magnetic type, respectively.

It is also amusing to verify that static configurations, for which
$\mathcal{N}=0$, are able to carry a nontrivial angular momentum,
because in the ``dyonic'' case the product of the electric and magnetic
$U(1)$ charges manifestly contribute to $\mathcal{J}$ in \eqref{eq:AngMom}.

Regularity of the configurations can also be analyzed from demanding
smoothness of the spacetime metric, so that $\mu_{\mathcal{P}}$ would
be related to the inverse of the Hawking temperature ($\mu_{\mathcal{P}}=-\beta$),
and $\mu_{\mathcal{J}}$ to the chemical potential associated to the
angular momentum $\mathcal{J}$, when it corresponds.

\subsection{Cosmological configurations and their thermodynamics}

In this subsection we discuss the class of stationary spherically
symmetric bosonic solutions for the case $\mathcal{P}-\frac{4\pi}{k}\mathcal{Z}^{2}\geq0$,
which clearly fulfills the energy bounds in section \ref{Bounds}.
The line element generically describes the class of locally flat cosmological
spacetimes discussed in \cite{Ezawa:1992nk}, \cite{Cornalba:2002fi},
\cite{Cornalba:2003kd} endowed with a generic choice of chemical
potentials as in \cite{Gary:2014ppa}, \cite{Matulich:2014hea}, and
reduces to the null orbifold in \cite{Horowitz:1990ap} for $\mathcal{P}=\frac{4\pi}{k}\mathcal{Z}^{2}$.
The thermodynamics of cosmological spacetimes in vacuum has been analyzed
in \cite{Barnich:2012xq}, \cite{Bagchi:2012xr}, \cite{Gary:2014ppa},
\cite{Matulich:2014hea}. Here we extend the analysis to the case
of cosmological spacetimes endowed with $U(1)$ charges of electric
and magnetic type. 

As explained in \cite{Matulich:2014hea} the Euclidean geometry possesses
the topology of a solid torus, so that the circles spanned by the
Euclidean time $\tau=-iu$, correspond to contractible cycles, and
the orientation is reversed as compared with the one of a BTZ black
hole \cite{Banados:1992wn}, \cite{Banados:1992gq}. Following refs.
\cite{Henneaux:2013dra} and \cite{Bunster:2014mua}, the solid torus
can be chosen to be parametrized as a ``straight'' one, i.e., described
by a fixed range of the coordinates, $0\leq\tau<1$ and $0\leq\phi<2\pi$.
In this way, the Hawking temperature and the chemical potential associated
to the angular momentum explicitly appear in the metric through $\mu_{\mathcal{P}}$
and $\mu_{\mathcal{J}}$, respectively. Thus, all of the chemical
potentials, including the ones for the $U(1)$ charges, given by $\mu_{\mathcal{Z}}$
and $\mu_{\mathcal{T}}$ become treated in the same footing. It is
worth pointing out that the chemical potentials of the $U(1)$ charges
have to be switched on, otherwise an interesting class of physical
solutions would fail to be regular.

Regularity of the configurations then implies that the holonomy of
the connection along a thermal circle has to be trivial. As explained
in section \ref{Regularity}, according to criterion $(i)$ we apply
a suitable gauge transformation given by $g=e^{\lambda_{0}M_{0}}$,
with $\lambda_{0}$ constant, so that the timelike component of the
connection \eqref{eq:au-bosonic} transforms as
\begin{eqnarray}
a_{u}^{g} & = & \mu_{\mathcal{J}}L_{1}-\frac{\pi}{k}\mu_{\mathcal{J}}\left(\mathcal{P}-\frac{4\pi}{k}\mathcal{Z}^{2}\right)L_{-1}+\left(\mu_{\mathcal{P}}+\lambda_{0}\mu_{\mathcal{J}}\right)M_{1}\nonumber \\
 &  & +\left(\mu_{\mathcal{T}}-\frac{2\pi}{k}\mu_{\mathcal{J}}\mathcal{Z}\right)T+\left(\mu_{\mathcal{Z}}-\frac{2\pi}{k}\mu_{\mathcal{J}}\mathcal{T}+\frac{8\pi}{k}\mu_{\mathcal{P}}\mathcal{Z}\right)Z\\
 &  & -\left[\frac{\pi}{k}\mu_{\mathcal{J}}\left(\mathcal{J}+\frac{2\pi}{k}\mathcal{T}\mathcal{Z}\right)+\frac{\pi}{k}\left(\mu_{\mathcal{P}}-\lambda_{0}\mu_{\mathcal{J}}\right)\left(\mathcal{P}-\frac{4\pi}{k}\mathcal{Z}^{2}\right)\right]M_{-1}\,.\nonumber 
\end{eqnarray}
Therefore, the components of $a_{u}^{g}$ along $M_{n}$ and $Z$
are gauged away for $\lambda_{0}=-\frac{\mu_{\mathcal{P}}}{\mu_{\mathcal{J}}}$,
provided that the chemical potentials of electric type, $\mu_{\mathcal{P}}$
and $\mu_{\mathcal{Z}}$, are fixed in terms of the magnetic one $\mu_{\mathcal{J}}$
and the global charges.
\begin{equation}
\mu_{\mathcal{P}}=-\frac{\mu_{\mathcal{J}}\left(\mathcal{J}+\frac{2\pi}{k}\mathcal{T}\mathcal{Z}\right)}{2\left(\mathcal{P}-\frac{4\pi}{k}\mathcal{Z}^{2}\right)}\quad,\quad\mu_{\mathcal{Z}}=\frac{2\pi}{k}\mu_{\mathcal{J}}\mathcal{T}-\frac{8\pi}{k}\mu_{\mathcal{P}}\mathcal{Z}\,.
\end{equation}
Note that the gauge transformation spanned by $g=e^{-\frac{\mu_{\mathcal{P}}}{\mu_{\mathcal{J}}}M_{0}}$
depends only on chemical potentials that are kept fixed at the boundary,
and hence it turns out to be a permissible one in the sense of \cite{Bunster:2014mua}.
The connection $a_{u}^{g}$ then reduces to the following one for
the $sl(2,R)\oplus u(1)$ algebra

\begin{equation}
a_{u}^{g}=\mu_{\mathcal{J}}\left[L_{1}-\frac{\pi}{k}\left(\mathcal{P}-\frac{4\pi}{k}\mathcal{Z}^{2}\right)L_{-1}\right]+\left(\mu_{\mathcal{T}}-\frac{2\pi}{k}\mu_{\mathcal{J}}\mathcal{Z}\right)T\,.\label{eq:agu}
\end{equation}
The remaining regularity condition $(ii)$ then corresponds to demanding
that \eqref{eq:agu} possesses a trivial holonomy, i.e., $H_{\mathcal{C}}=\pm\mathbb{I}_{2\times2}$,
which can be readily performed in the fundamental representation of
$sl(2,R)\oplus u(1)$ (see appendix \ref{Conventions}). Therefore,
the eigenvalues of $a_{\tau}^{g}=ia_{u}^{g}$, given by
\begin{equation}
\Lambda_{[\pm]}=\frac{1}{2}\left(\text{tr}\left[\hat{\omega}\right]\pm\sqrt{2\text{tr}\left[\hat{\omega}^{2}\right]-\text{tr}\left[\hat{\omega}\right]^{2}}\right)\,,\label{eq:Lambdapm}
\end{equation}
with $\hat{\omega}=a_{\tau}^{g}$, have to be given by $\Lambda_{[\pm]}=\pm i\pi m$,
with $m$ an arbitrary integer (even for $\mathbb{I}_{2\times2}$,
and odd for $-\mathbb{I}_{2\times2}$). The regularity condition then
reduces to
\begin{equation}
\pm i\pi m=-\left(\mu_{\mathcal{T}}-\frac{2\pi}{k}\mu_{\mathcal{J}}\mathcal{Z}\right)\pm i\sqrt{\frac{\pi}{k}}|\mu_{\mathcal{J}}|\left(\mathcal{P}-\frac{4\pi}{k}\mathcal{Z}^{2}\right)^{1/2},
\end{equation}
which implies that
\begin{equation}
\mu_{\mathcal{T}}=\frac{2\pi}{k}\mu_{\mathcal{J}}\mathcal{Z}\quad,\quad|\mu_{\mathcal{J}}|=\frac{m\sqrt{\pi k}}{\left(\mathcal{P}-\frac{4\pi}{k}\mathcal{Z}^{2}\right)^{1/2}}\,.
\end{equation}
In sum, regularity implies that the chemical potentials become fixed
in terms of the global charges according to
\begin{eqnarray}
\mu_{\mathcal{P}} & = & -\frac{m\sqrt{\pi k}|\mathcal{J}+\frac{2\pi}{k}\mathcal{T}\mathcal{Z}|}{2\left(\mathcal{P}-\frac{4\pi}{k}\mathcal{Z}^{2}\right)^{3/2}}\,,\nonumber \\
\mu_{\mathcal{J}} & = & \text{sgn}\left(\mathcal{J}+\frac{2\pi}{k}\mathcal{T}\mathcal{Z}\right)\frac{m\sqrt{\pi k}}{\left(\mathcal{P}-\frac{4\pi}{k}\mathcal{Z}^{2}\right)^{1/2}}\,,\nonumber \\
\mu_{\mathcal{T}} & = & \text{sgn}\left(\mathcal{J}+\frac{2\pi}{k}\mathcal{T}\mathcal{Z}\right)\frac{2m\pi^{3/2}\mathcal{Z}}{\sqrt{k}\left(\mathcal{P}-\frac{4\pi}{k}\mathcal{Z}^{2}\right)^{1/2}}\,,\label{eq:musreg}\\
\mu_{\mathcal{Z}} & = & \text{sgn}\left(\mathcal{J}+\frac{2\pi}{k}\mathcal{T}\mathcal{Z}\right)\frac{2m\pi^{3/2}\mathcal{T}}{\sqrt{k}\left(\mathcal{P}-\frac{4\pi}{k}\mathcal{Z}^{2}\right)^{1/2}}+\frac{4m\pi^{3/2}|\mathcal{J}+\frac{2\pi}{k}\mathcal{T}\mathcal{Z}|\mathcal{Z}}{\sqrt{k}\left(\mathcal{P}-\frac{4\pi}{k}\mathcal{Z}^{2}\right)^{3/2}}\,.\nonumber 
\end{eqnarray}
It is worth noting that the branch that is connected to the standard
cosmological spacetime, so that the Hawking temperature is given by
$\mu_{\mathcal{P}}=-\beta$, corresponds to $m=1$, which agrees with
what is found from requiring smoothness of the Euclidean metric.

Since we are dealing with a Chern-Simons theory, the entropy associated
to the cosmological horizon can be directly found from the following
formula \cite{Perez:2012cf}, \cite{Perez:2013xi}, \cite{deBoer:2013gz},
\cite{Bunster:2014mua}
\begin{equation}
S=\frac{k}{2\pi}\int\langle a_{\tau}a_{\phi}\rangle d\phi\,,
\end{equation}
which by virtue of \eqref{eq:aphi-bosonic} and \eqref{eq:au-bosonic},
evaluates as

\begin{equation}
S=2\pi\left(2\mu_{\mathcal{J}}\mathcal{J}+2\mu_{\mathcal{P}}\mathcal{P}+\mu_{\mathcal{T}}\mathcal{T}+\mu_{\mathcal{Z}}\mathcal{Z}\right)\,.
\end{equation}
Therefore, as required by the action principle, the entropy of regular
configurations, by virtue of \eqref{eq:musreg} reduces to
\begin{equation}
S=2\pi\left[m\sqrt{\frac{\pi k}{\mathcal{P}-\frac{4\pi}{k}\mathcal{Z}^{2}}}|\mathcal{J}+\frac{2\pi}{k}\mathcal{T}\mathcal{Z}|\right]\,.\label{eq:S1}
\end{equation}
Note that for $m=1$, the entropy
\begin{equation}
S=2\pi\sqrt{\frac{\pi k}{\mathcal{P}-\frac{4\pi}{k}\mathcal{Z}^{2}}}|\mathcal{J}+\frac{2\pi}{k}\mathcal{T}\mathcal{Z}|\,,\label{eq:Entropy}
\end{equation}
precisely gives a quarter of the cosmological horizon area over $4G$. 

It is amusing to verify that, unlike the case of pure gravity, configurations
without angular momentum ($\mathcal{J}=0$) can still be stationary
and carry a nonvanishing entropy.

It is also worth pointing out that the entropy expressed in terms
of the extensive variables, manifestly depends not only on the energy
and the angular momentum, but also on the electric and magnetic $U(1)$
charges. Indeed, it is simple to verify that the first law holds in
the grand canonical ensemble, since 
\begin{equation}
\delta S=2\pi\left(\mu_{\mathcal{P}}\delta\mathcal{P}+\mu_{\mathcal{J}}\delta\mathcal{J}+\mu_{\mathcal{T}}\delta\mathcal{T}+\mu_{\mathcal{Z}}\delta\mathcal{Z}\right)\,,
\end{equation}
provided that the chemical potentials are given by \eqref{eq:musreg}.
In other words, an attempt of fixing the chemical potentials in a
different way as in eq. \eqref{eq:musreg}, might generate a severe
clash with the first law of thermodynamics.

As an ending remark of this section, it is worth mentioning that the
entropy of the class of cosmological spacetimes with $U(1)$ fields
in \eqref{eq:Entropy}, has also been obtained simultaneously and
in an independent way through a different approach in \cite{Basu:2017aqn},
which is certainly reassuring.

\subsection{Conical defects with $U(1)$ fluxes}

In the case of $-\frac{k}{4\pi}<\mathcal{P}-\frac{4\pi}{k}\mathcal{Z}^{2}<0$,
the configurations are described by spacetime metrics \eqref{eq:Line-Element}
that generically describe rotating conical defects \cite{Deser:1983tn},
\cite{Deser:1983nh} with nontrivial lapse and shift functions, endowed
with $U(1)$ fields of electric and magnetic type that do not generate
a back reaction. In the case of antiperiodic boundary conditions the
energy bounds in section \ref{Bounds} are clearly satisfied, but
this is not necessarily so for the case of periodic boundary conditions.
Indeed, for periodic boundary conditions, if $|\mathcal{Z}|\geq\frac{k}{4\pi}$,
the stronger energy bound is trivially satisfied and never saturates;
while for $|\mathcal{Z}|<\frac{k}{4\pi}$, the bound is fulfilled
only for configurations with nonnegative energy ($\mathcal{P}\geq0$).
Note that in vacuum ($\mathcal{Z}=\mathcal{T}=0$), this class of
configurations does not fulfill the energy bounds when the spin structure
is even.

Although this class of configurations fails to be regular due to the
presence of sources at the origin, it turns out to be very interesting.
This is because they might admit Killing spinors provided that the
$U(1)$ fluxes are suitably tuned with the angular deficit \cite{Howe:1995zm},
\cite{Edelstein:1995md}. In our terms, these $U(1)$ fluxes correspond
to charges of electric type. In section \ref{Bosonic}, the explicit
form of the Killing spinors is constructed, and we also show that
two of our supersymmetry bounds in section \ref{Bounds} are saturated,
so that the configurations correspond to half-BPS states. 

\subsection{Minkowski spacetime and conical surpluses endowed with $U(1)$ fields
\label{Surpluses}}

Configurations with $\mathcal{P}-\frac{4\pi}{k}\mathcal{Z}^{2}<-\frac{k}{4\pi}$
correspond to spacetime metrics with conical surpluses, so that they
possess angular excess. Despite they do not fulfill the energy bounds,
they might be interesting because it can be seen that they could also
admit Killing spinors provided that suitable $U(1)$ fluxes are switched
on. Nonetheless, it should be emphasized that they cannot describe
BPS states. 

In the case of $\mathcal{P}-\frac{4\pi}{k}\mathcal{Z}^{2}=-\frac{k}{4\pi}$
the bosonic solution is described by the Minkowski spacetime endowed
with electric and magnetic $U(1)$ fields. Noteworthy, the energy
bounds for the antiperiodic boundary conditions are clearly saturated,
but this is not necessarily so for periodic boundary conditions. 

Remarkably, in the absence of angular momentum and magnetic $U(1)$
charges, these configurations can be regarded as regular ones provided
that the energy and the electric $U(1)$ charges take discrete values,
according to

\begin{equation}
\mathcal{P}=\frac{k}{\pi}n_{+}n_{-}\quad,\quad\mathcal{Z}=-\frac{k}{4\pi}\left(n_{+}+n_{-}\right)\,,\label{eq:electricquant}
\end{equation}
where $n_{\pm}$ correspond to (half-)integers in the case of (anti)periodic
boundary conditions. Note that for antiperiodic boundary conditions,
the electric $U(1)$ charge is then given by an integer multiple of
the mass of Minkowski spacetime in vacuum ($\mathcal{P}=-\frac{k}{4\pi}$),
while for periodic boundary conditions, it corresponds to an even
multiple.

This can be seen as follows. In these cases, regularity of the configurations
corresponds to requiring the holonomy of the connection for a contractible
circle along the angular coordinate to be trivial. Thus, following
the criterion $(i)$ in section \ref{Regularity}, one can show that
there is no group element of the form $g=e^{\lambda_{n}M_{n}+\lambda Z}$
that helps in order to gauge away the components of $a_{\phi}$ in
\eqref{eq:aphi-bosonic} along the generators $M_{n}$ and $Z$, and
hence regularity necessarily implies that the angular momentum and
the magnetic $U(1)$ charge vanish, i.e., $\mathcal{J}=\mathcal{T}=0$.

The connection $a_{\phi}$ then reduces to
\begin{equation}
a_{\phi}=L_{1}-\frac{\pi}{k}\left(\mathcal{P}-\frac{4\pi}{k}\mathcal{Z}^{2}\right)L_{-1}-\frac{2\pi}{k}\mathcal{Z}T\,,\label{eq:aphi-sur}
\end{equation}
that takes values on the $sl(2,R)\oplus u(1)$ subalgebra. Criterion
$(ii)$ then implies that the holonomy of $a_{\phi}$ in \eqref{eq:aphi-sur}
along the angular circle is trivial ($H_{\mathcal{C}}=\pm\mathbb{I}_{2\times2}$).
Hence, the eigenvalues of $\hat{\omega}=a_{\phi}$, can be obtained
from \eqref{eq:Lambdapm}, and they have to be given by $\Lambda_{[\pm]}=i\lambda_{[\pm]}=in_{\pm}$,
where $n_{\pm}$ stand for integers or half-integers in the cases
of even or odd spin structures, respectively, and $\lambda_{[\pm]}$
are given in eq. \eqref{eq:lambdapm}. The regularity condition then
reads
\begin{equation}
n_{\pm}=-\frac{2\pi}{k}\mathcal{Z}\pm\sqrt{\frac{\pi}{k}\left(\frac{4\pi}{k}\mathcal{Z}^{2}-\mathcal{P}\right)}\,,
\end{equation}
which implies that the energy and the electric $U(1)$ charge take
discrete values given by \eqref{eq:electricquant}.

In particular, it is worth pointing out that the configurations that
correspond to Minkowski spacetime endowed with electric $U(1)$ charge
turn out to be regular for $n_{+}=n_{-}+1$, which implies that
\begin{equation}
\mathcal{Z}=-\frac{k}{4\pi}\left(2n_{-}+1\right)\,,\label{eq:Zquant}
\end{equation}
is given by even or odd multiples of the mass of Minkowski spacetime
in vacuum for antiperiodic or periodic boundary conditions, respectively;
while the total energy of the configuration is given by

\begin{equation}
\mathcal{P}=\frac{k}{\pi}n_{-}\left(n_{-}+1\right)\,.\label{eq:Pquant}
\end{equation}
Remarkably, Minkowski spacetime endowed with an electric $U(1)$ charge
given by \eqref{eq:Zquant} fulfills all of the energy bounds and
saturates four of them for periodic or antiperiodic boundary conditions
for the fermions, so that the configurations turn out to be maximally
supersymmetric, admitting four Killing spinors whose explicit form
is given in the next section. This is in stark contrast with what
occurs for Minkowski spacetime in vacuum, which does not fulfill the
energy bounds in the case of periodic boundary conditions for the
fermions. In other words, for an even spin structure, Minkowski spacetime
can be brought back into the allowed spectrum provided that it is
endowed with an electric-like $U(1)$ charge given by an odd multiple
of the mass of the Minkowski spacetime in vacuum.

As a closing remark of this subsection, it is worth pointing out that
similar classes of solitonic-like objects with conical surpluses have
also been disscused in the context of three-dimensional gravity coupled
to higher spin fields in refs. \cite{Chen:1303}, \cite{Henneaux:2015ywa},
\cite{Castro:1111}, \cite{Datta:1208}, \cite{Campoleoni:1307},
\cite{Campoleoni:1307-1}, \cite{Li:1308}, \cite{Raeymaekers:1412},
\cite{Fuentealba:2015wza}.

\section{Bosonic solutions with unbroken supersymmetries\label{Bosonic}}

\subsection{Asymptotic Killing spinor equations \label{AsymptKS}}

Here we look for the class of bosonic configurations that asymptotically
behave as the stationary spherically symmetric ones described by the
gauge fields in \eqref{eq:aphi-bosonic} and \eqref{eq:au-bosonic},
that admit Killing spinors being well-defined in the asymptotic region.
These unbroken supersymmetries are spanned by fermionic gauge transformations
that leave the bosonic configuration invariant. Thus, they fulfill
$\delta a=d\lambda+[a,\lambda]=0$, where $\lambda$ stands for the
purely fermionic asymptotic symmetries that can be obtained from eqs.
\eqref{eq:Lambda} and \eqref{eq:Lambachico}. The fermionic Lie-algebra-valued
parameter then reads
\begin{equation}
\lambda\left[\epsilon_{\mathcal{\psi}},\epsilon_{\mathcal{S}}\right]=\epsilon_{\psi}G_{\frac{1}{2}}^{1}-\left(\epsilon_{\mathcal{\psi}}\text{\ensuremath{'}}-\frac{2\pi}{k}\mathcal{Z}\epsilon_{\mathcal{S}}\right)G_{-\frac{1}{2}}^{1}+\epsilon_{S}G_{\frac{1}{2}}^{2}-\left(\epsilon_{\mathcal{S}}\text{\ensuremath{'}}+\frac{2\pi}{k}\mathcal{Z}\epsilon_{\psi}\right)G_{-\frac{1}{2}}^{2}=\epsilon_{I}^{\alpha}Q_{\alpha}^{I}\,,\label{eq:lambdaferm}
\end{equation}
so that the explicit form of the spinors $\epsilon_{I}^{\alpha}$
in terms of the Grasmann-valued functions $\epsilon_{\mathcal{\psi}},\epsilon_{\mathcal{S}}$,
can then be obtained by virtue of the change of basis in \eqref{eq:ChangeBasis}.
The spinors are then given by
\begin{equation}
\epsilon_{1}=\sqrt{2}\left(\begin{array}{c}
-\epsilon_{\mathcal{\psi}}\text{\ensuremath{'}}+\frac{2\pi}{k}\mathcal{Z}\epsilon_{\mathcal{S}}\\
\epsilon_{\psi}
\end{array}\right)\quad,\quad\epsilon_{2}=\sqrt{2}\left(\begin{array}{c}
-\epsilon_{\mathcal{S}}\text{\ensuremath{'}}-\frac{2\pi}{k}\mathcal{Z}\epsilon_{\psi}\\
\epsilon_{S}
\end{array}\right)\,.\label{eq:EPSILONS}
\end{equation}

The asymptotic Killing spinor equations can be partially read from
the transformation law of the fields in \eqref{eq:delta} under the
asymptotic supersymmetries. The nontrivial ones are given by
\begin{eqnarray}
\delta\psi & = & -\mathcal{P}\epsilon_{\psi}+\frac{k}{\pi}\epsilon_{\psi}\text{\ensuremath{''}}-4\mathcal{Z}\epsilon_{\mathcal{S}}\text{\ensuremath{'}}=0\,,\nonumber \\
\delta\mathcal{S} & = & -\mathcal{P}\epsilon_{\mathcal{S}}+\frac{k}{\pi}\epsilon_{\mathcal{S}}\text{\ensuremath{''}}+4\mathcal{Z}\epsilon_{\psi}\text{\ensuremath{'}}=0\,.\label{eq:deltafermionic}
\end{eqnarray}
Analogously, the conditions for the Grasmann-valued parameters can
be obtained from \eqref{eq:Chiralities}, so that they read
\begin{eqnarray}
\dot{\epsilon}_{\psi} & = & \mu_{\mathcal{J}}\epsilon_{\psi}\text{\ensuremath{'}}-\mu_{\mathcal{T}}\epsilon_{\mathcal{S}}\,,\nonumber \\
\dot{\epsilon}_{\mathcal{S}} & = & \mu_{\mathcal{J}}\epsilon_{\mathcal{S}}\text{\ensuremath{'}}+\mu_{\mathcal{T}}\epsilon_{\psi}\,.\label{eq:puntofermionic}
\end{eqnarray}
In order to perform the analysis it is useful to define a single complex
Grasmann-valued parameter defined as 
\begin{equation}
\xi=\frac{1}{\sqrt{2}}\left(\epsilon_{\psi}+i\epsilon_{\mathcal{S}}\right)\,,
\end{equation}
so that the asymptotic Killing spinor equations in \eqref{eq:deltafermionic}
and \eqref{eq:puntofermionic} can be rewritten as
\begin{equation}
\xi\text{\ensuremath{''}}+\frac{4\pi}{k}i\mathcal{Z}\xi\text{\ensuremath{'}}-\frac{\pi}{k}\mathcal{P}\xi=0\,,\label{eq:AKsE}
\end{equation}
\begin{equation}
\dot{\xi}-\mu_{\mathcal{J}}\xi\text{\ensuremath{'}}-i\mu_{\mathcal{T}}\xi=0\,.\label{eq:chiE}
\end{equation}
The solution of the asymptotic Killing spinor equations in \eqref{eq:AKsE}
and \eqref{eq:chiE} can then readily found. In the generic case $\left(\mathcal{P}\neq\frac{4\pi}{k}\mathcal{Z}^{2}\right)$,
the solution reads
\begin{equation}
\xi=\xi_{1}e^{i\mu_{\mathcal{T}}u}e^{i\lambda_{[+]}\hat{\phi}}+\xi_{2}e^{i\mu_{\mathcal{T}}u}e^{i\lambda_{[-]}\hat{\phi}}\,,\label{eq:xi-2}
\end{equation}
where $\lambda_{[\pm]}$ correspond to the eigenvalues of the spacelike
components of the $sl(2,R)\oplus u(1)$ connection $\hat{\omega}=\omega+B$
in \eqref{eq:lambdapm}, and $\hat{\phi}:=\phi+\mu_{\mathcal{J}}u$.
Here $\xi_{1}$ and $\xi_{2}$ stand for arbitrary complex (Grasmann-valued)
constants. In the special case of $\mathcal{P}=\frac{4\pi}{k}\mathcal{Z}^{2}$,
which corresponds to energy of the null orbifold endowed with $U(1)$
fields, the eigenvalues degenerate ($\lambda_{[+]}=\lambda_{[-]}=-\frac{2\pi}{k}\mathcal{Z}$),
and hence the solution is given by
\begin{equation}
\xi=\xi_{1}e^{i\mu_{\mathcal{T}}u}e^{i\lambda_{[+]}\hat{\phi}}+\xi_{2}\phi e^{\left(\mu_{\mathcal{J}}+i\mu_{\mathcal{T}}\right)u}e^{i\lambda_{[+]}\hat{\phi}}\,.\label{eq:DegSol}
\end{equation}

One is then ready to analyze whether they are well-defined for the
different classes of bosonic solutions discussed in section \ref{Configurations}. 

\subsubsection{Cosmological configurations}

In this case, the energy fulfills $\mathcal{P}>\frac{4\pi}{k}\mathcal{Z}^{2}$,
so that the eigenvalues $\lambda_{[\pm]}$ in \eqref{eq:lambdapm}
turn out to be complex. Therefore, the solution for the Killing spinor
equations in \eqref{eq:xi-2} is not globally well-defined because
it cannot fulfill neither periodic nor antiperiodic boundary conditions.
Consequently, all of the supersymmetries are broken, which goes by
hand with the fact the energy bounds in this case are always satisfied,
but never saturated.

\subsubsection{Null orbifold with $U(1)$ fields}

For this class of configurations, the energy is given by $\mathcal{P}=\frac{4\pi}{k}\mathcal{Z}^{2}$,
so that the solution of the asymptotic Killing spinor equations is
given by \eqref{eq:DegSol}. It is then clear that it can only be
globally defined provided that $\xi_{2}=0$, where the eigenvalue
has to be given by $\lambda_{[+]}=n$, with $n$ a (half-)integer
for fermions that fulfill (anti)periodic boundary conditions. Note
that the electric $U(1)$ charge is then restricted to take the following
values: $\mathcal{Z}=-\frac{kn}{2\pi}$. The solution of the asymptotic
Killing spinor equation then acquires the form
\begin{equation}
\xi=\xi_{1}e^{i\mu_{\mathcal{T}}u}e^{-in\hat{\phi}}\,,
\end{equation}
so that it is clear that this class of configurations possesses two
unbroken supersymmetries. This is precisely the number of bounds that
saturate, which correspond to the ones in \eqref{eq:Bpmr} for $r=\mp\lambda_{[+]}=\mp n=\pm\frac{2\pi}{k}\mathcal{Z}$.

\subsubsection{Conical defects with $U(1)$ fluxes}

In this case the energy fulfills the condition $-\frac{k}{4\pi}<\mathcal{P}-\frac{4\pi}{k}\mathcal{Z}^{2}<0$,
which in terms of the eigenvalues $\lambda_{[\pm]}$ in \eqref{eq:lambdapm},
reads
\begin{equation}
\lambda_{[+]}-\lambda_{[-]}<1\,.
\end{equation}
Therefore, the solution of the asymptotic Killing spinor equation
in \eqref{eq:xi-2} is globally well-defined provided that $\xi_{2}=0$
and $\lambda_{[+]}=n_{+}$ is given by a (half-)integer, or $\xi_{1}=0$
and $\lambda_{[-]}=n_{-}$ a (half-)integer, for (anti)periodic boundary
conditions. Therefore, this class of configurations preserves half
of the supersymmetries, which goes by hand with the fact that the
energy bounds are fulfilled, and the number of them that saturate
is just two. They correspond to the ones in \eqref{eq:Bpmr} either
for $r=\mp\lambda_{[+]}$, or $r=\mp\lambda_{[-]}$.

It is worth then mentioning that configurations with conical defects
become half-BPS for even or odd spin structures, provided that the
electric $U(1)$ charge and the energy are suitably tuned according
to
\begin{equation}
\mathcal{P}=\frac{k}{\pi}n_{+}n_{-}\quad,\quad\mathcal{Z}=-\frac{k}{4\pi}\left(n_{+}+n_{-}\right)\,.
\end{equation}

\subsubsection{Minkowski spacetime with $U(1)$ fields}

In the case of Minkowski spacetime dressed with $U(1)$ fields, the
energy is given by $\mathcal{P}=-\frac{k}{4\pi}+\frac{4\pi}{k}\mathcal{Z}^{2}$,
which implies that the eigenvalues $\lambda_{[\pm]}$ in \eqref{eq:lambdapm}
fulfill
\begin{equation}
\lambda_{[+]}-\lambda_{[-]}=1\,.
\end{equation}
Therefore, the asymptotic Killing spinors are given by \eqref{eq:xi-2},
and they are globally well-defined provided that $\lambda_{[\pm]}$
are given by (half-)integers for (anti)periodic boundary conditions.
This is the only maximally supersymmetric case that fulfill the bounds
in section \ref{Bounds}, which agrees with the fact that four of
them in \eqref{eq:Bpmr} saturate, corresponding to the cases of $r=\mp(\lambda_{[-]}+1)$
and $r=\mp\lambda_{[-]}$.

Note that the energy and the electric $U(1)$ charge are then given
by \eqref{eq:Pquant} and \eqref{eq:Zquant}, respectively, with $\lambda_{[-]}=n_{-}$.

It is worth emphasizing that this case includes the regular one described
in section \ref{Surpluses}, which corresponds to configurations with
$\mathcal{J}=\mathcal{T}=0$. 

\subsubsection{Configurations with conical surpluses }

For this class of configurations the energy lies in the range $\mathcal{P}<-\frac{k}{4\pi}+\frac{4\pi}{k}\mathcal{Z}^{2}$,
which amounts to
\begin{equation}
\lambda_{[+]}-\lambda_{[-]}>1\,.
\end{equation}
As aforementioned, the energy bounds are not fulfilled, but nonetheless
this class of solutions might admit unbroken supersymmetries. In the
maximally supersymmetric case the asymptotic Killing spinors are given
by \eqref{eq:xi-2} provided that $\lambda_{[\pm]}$ correspond to
(half-)integers for (anti)periodic boundary conditions. Note that
in the case of $\mathcal{J}=\mathcal{T}=0$, this class of configurations
turns out to be regular (see section \ref{Surpluses}). The global
charges are then given by \eqref{eq:electricquant}, with $\lambda_{[\pm]}=n_{\pm}$.

The remaining possibility consists on configurations with angular
excess endowed with suitable electric $U(1)$ charges, which are not
regular. The asymptotic Killing spinors are then given by \eqref{eq:xi-2},
and they are globally well-defined either for $\xi_{2}=0$ and a (half-)integer
$\lambda_{[+]}=n_{+}$, or $\xi_{1}=0$ and $\lambda_{[-]}=n_{-}$
a (half-)integer, for (anti)periodic boundary conditions. 

\subsection{Global Killing spinor equation \label{GlobalKilling}}

For the class of exact solutions described in section \ref{Configurations},
one can also proceed in the standard way in order to identify the
configurations that admit globally well-defined Killing spinors, as
well as finding the explicit form of them. It is reassuring to verify
that the results in this section precisely agree with the ones in
section \ref{AsymptKS}. Indeed, the results within this section can
be readily reconstructed from the ones in section \ref{AsymptKS}
by virtue of eq. \eqref{eq:EPSILONS}. Nonetheless, carrying out the
analysis in the standard way turns out to be a healthy exercise. In
order to perform this task it is useful to complexify the spinors
according to
\begin{equation}
\chi=\frac{1}{\sqrt{2}}\left(\epsilon_{1}+i\epsilon_{2}\right)\,,
\end{equation}
so that the Killing spinor equation can be obtained from the local
supersymmetry transformations in \eqref{eq:localsusy} ($\delta\psi_{I}=0$),
and it turns out to be given by
\begin{equation}
\delta\chi=d\chi+\frac{1}{2}\omega^{a}\Gamma_{a}\chi-iB\chi=0\,.\label{eq:complexKSE}
\end{equation}
The generic solution of \eqref{eq:complexKSE} is given by
\begin{equation}
\chi=\text{\ensuremath{\mathcal{P}}exp}\left[-\int_{\mathcal{C}}\left(\frac{1}{2}\omega^{a}\Gamma_{a}-iB\right)\right]\chi_{0}\,,\label{eq:ComplexKS}
\end{equation}
with $\chi_{0}$ a constant Dirac spinor. 

As it can be read from section \ref{Configurations}, the spin connection
and the electric $U(1)$ field read
\begin{eqnarray}
\omega & = & \sqrt{2}\left[J_{1}+\frac{\pi}{k}\left(\mathcal{P}-\frac{4\pi}{k}\mathcal{Z}^{2}\right)J_{0}\right]\left(d\phi+\mu_{\mathcal{J}}du\right)\,,\\
B & = & -\frac{2\pi}{k}\mathcal{Z}\left(d\phi+\mu_{\mathcal{J}}du\right)+\mu_{\mathcal{T}}du\,,
\end{eqnarray}
and hence, the solution in \eqref{eq:ComplexKS} reduces to
\begin{equation}
\chi=e^{i\left(-\frac{2\pi}{k}\mathcal{Z}\hat{\phi}+\mu_{\mathcal{T}}u\right)}\text{exp}\left\{ -\sqrt{2}\left[J_{1}+\frac{\pi}{k}\left(\mathcal{P}-\frac{4\pi}{k}\mathcal{Z}^{2}\right)J_{0}\right]\hat{\phi}\right\} \chi_{0}\,,\label{eq:KS1}
\end{equation}
with $\hat{\phi}=\phi+\mu_{\mathcal{J}}u$.

In the generic case of ($\mathcal{P}\neq\frac{4\pi}{k}\mathcal{Z}^{2}$),
the solution in \eqref{eq:KS1} then acquires the form
\begin{eqnarray}
\chi & = & e^{i\left(-\frac{2\pi}{k}\mathcal{Z}\hat{\phi}+\mu_{\mathcal{T}}u\right)}\left\{ \cosh\left[-\frac{i}{4}(\lambda_{[+]}-\lambda_{[-]})\hat{\phi}\right]\mathbb{I}_{2\times2}\right.\nonumber \\
 &  & \left.-\frac{4i\sqrt{2}}{\lambda_{[+]}-\lambda_{[-]}}\left[J_{1}-\frac{1}{16}(\lambda_{[+]}-\lambda_{[-]})^{2}J_{0}\right]\sinh\left[-\frac{i}{4}(\lambda_{[+]}-\lambda_{[-]})\hat{\phi}\right]\right\} \chi_{0\,,}\label{eq:KS2}
\end{eqnarray}
with $\lambda_{[\pm]}$ defined in eq. \eqref{eq:lambdapm}. 

For $\mathcal{P}=\frac{4\pi}{k}\mathcal{Z}^{2}$, the solution is
given by
\begin{equation}
\chi=\left(\mathbb{I}_{2\times2}-\sqrt{2}J_{1}\hat{\phi}\right)e^{i\left(-\frac{2\pi}{k}\mathcal{Z}\hat{\phi}+\mu_{\mathcal{T}}u\right)}\chi_{0}\,.\label{eq:KS3}
\end{equation}
Here we have made use of the fundamental matrix representation of
$sl(2,R)\oplus u(1)$ that appears in appendix \ref{Conventions}.

The remaining analysis can then be directly performed as in the previous
section.

In the case of cosmological configurations ($\mathcal{P}>\frac{4\pi}{k}\mathcal{Z}^{2}$)
all of the supersymmetries are broken because the Killing spinors
in \eqref{eq:KS2} are clearly not globally well-defined.

The null orbifold with $U(1)$ fields ($\mathcal{P}=\frac{4\pi}{k}\mathcal{Z}^{2}$)
possesses half of the supersymmetries provided that $\mathcal{Z}=\frac{kn}{2\pi}$,
so that the Killing spinors can be obtained from \eqref{eq:KS3}.
They are explicitly given by
\begin{equation}
\chi=e^{i\left(-n\hat{\phi}+\mu_{\mathcal{T}}u\right)}\chi_{0}\,,
\end{equation}
where the constant spinor has to fulfill the projection $J_{1}\chi_{0}=0$,
and $n$ is a (half-)integer for (anti)periodic boundary conditions.

In the case of conical defects with $U(1)$ fluxes ($-\frac{k}{4\pi}<\mathcal{P}-\frac{4\pi}{k}\mathcal{Z}^{2}<0$),
the Killing spinor in \eqref{eq:KS2} becomes
\begin{eqnarray}
\chi & = & e^{i\left(-\frac{2\pi}{k}\mathcal{Z}\hat{\phi}+\mu_{\mathcal{T}}u\right)}\left\{ \cos\left[\frac{1}{4}(\lambda_{[+]}-\lambda_{[-]})\hat{\phi}\right]\mathbb{I}_{2\times2}\right.\nonumber \\
 &  & \left.-\frac{4\sqrt{2}}{\lambda_{[+]}-\lambda_{[-]}}\left[J_{1}-\frac{1}{16}(\lambda_{[+]}-\lambda_{[-]})^{2}J_{0}\right]\sin\left[\frac{1}{4}(\lambda_{[+]}-\lambda_{[-]})\hat{\phi}\right]\right\} \chi_{0}\,,\nonumber \\
\label{eq:Chi}
\end{eqnarray}
which is globally well-defined provided that the spinor $\chi_{0}$
satisfies the projection
\begin{equation}
-\frac{4\sqrt{2}}{\lambda_{[+]}-\lambda_{[-]}}\left[J_{1}-\frac{1}{16}(\lambda_{[+]}-\lambda_{[-]})^{2}J_{0}\right]\chi_{0}=\pm i\chi_{0}\,.\label{eq:SelfCond}
\end{equation}
The Killing spinor in \eqref{eq:Chi} then reduces to
\begin{equation}
\chi=e^{i\mu_{\mathcal{T}}u}\exp\left\{ i\lambda_{[\pm]}\hat{\phi}\right\} \chi_{0}\,.\label{eq:chiconico}
\end{equation}
with $\lambda_{[\pm]}$ given by \eqref{eq:lambdapm}, and the sign
of the projection in \eqref{eq:SelfCond} coincides with the choice
of the label of $\lambda_{[\pm]}$ in \eqref{eq:chiconico}. The spinors
are then consistent with (anti)periodic boundary conditions provided
that $\lambda_{\pm}$ is a (half-)integer. Note that the projection
condition in \eqref{eq:SelfCond}, preserves half of the supersymmetries.

In the case of Minkowski spacetime dressed with $U(1)$ fields ($\mathcal{P}=-\frac{k}{4\pi}+\frac{4\pi}{k}\mathcal{Z}^{2}$),
equation \eqref{eq:KS2} reduces to

\begin{equation}
\chi=\left[\cos\left(\frac{\hat{\phi}}{2}\right)\mathbb{I}_{2\times2}-2\sqrt{2}\left(J_{1}-\frac{1}{4}J_{0}\right)\sin\left(\frac{\hat{\phi}}{2}\right)\right]e^{i\left(-\frac{2\pi}{k}\mathcal{Z}\hat{\phi}+\mu_{\mathcal{T}}u\right)}\chi_{0}\,,
\end{equation}
so that the Killing spinors are globally well-defined provided the
electric $U(1)$ charge is fixed by even or odd multiples of $-\frac{k}{4\pi}$
for antiperiodic or periodic boundary conditions, respectively. Noteworthy,
this case is maximally supersymmetric for even and odd spin structures.

For the class of conical surpluses endowed with $U(1)$ fluxes ($\mathcal{P}<-\frac{k}{4\pi}+\frac{4\pi}{k}\mathcal{Z}^{2}$)
there are two possibilities. 

In the maximally supersymmetric case the Killing spinors are given
by \eqref{eq:Chi}, where $\lambda_{[\pm]}=n_{\pm}$ stand for (half-)integers
in the case of (anti)periodic boundary conditions. 

In the remaining possibility the configurations possess two independent
Killing spinors, now given by \eqref{eq:chiconico}, so that $\chi_{0}$
fulfills the projection in \eqref{eq:SelfCond}. The Killing spinors
fulfill (anti)periodic boundary conditions when $\lambda_{\pm}$ is
a (half-)integer. 

\section{Spectral flow: from Ramond to Neveu-Schwarz boundary conditions \label{Spectral}}

As shown in \cite{Schwimmer:1986mf}, the spectrum spanned by the
super-Virasoro algebra with $\mathcal{N}=2$ in the case of periodic
boundary conditions for the fermionic fields relates to the one for
antiperiodic boundary conditions through spectral flow. These results
were generalized for $\mathcal{N}>2$ in \cite{Henneaux:1999ib}. 

Here we show that this is also the case for the super-BMS$_{3}$ algebra
with $\mathcal{N}=(2,0)$ in eq. \eqref{eq:alg3}. This occurs by
virtue of a field redefinition that is induced by a suitable $U(1)$
gauge transformation. The field redefinition then amounts to a precise
change of basis in the canonical generators that turns out to be an
automorphism of the super-BMS$_{3}$ algebra with $\mathcal{N}=(2,0)$.
This can be seen as follows.

The Dirac spinors $\xi^{\pm}=\frac{1}{\sqrt{2}}\left(\psi_{1}\pm i\psi_{2}\right)$
are such that $\xi^{+}$ is the hermitian conjugate of $\xi^{-}$,
and vice versa. Therefore, for generic ``anyonic'' boundary conditions
characterized by some parameter $\eta$ that is held fixed at the
boundary, the spinors fulfill
\begin{equation}
\xi^{\pm}\left(\phi+2\pi\right)=e^{\pm2i\pi\eta}\xi^{\pm}\left(\phi\right)\,,\label{eq:Anyonic}
\end{equation}
where $\eta=0,\frac{1}{2}$ corresponds to Ramond (periodic), and
Neveu-Schwarz (antiperiodic) boundary conditions, respectively. 

Besides, under $U(1)$ gauge transformation spanned by $g=e^{fT}$,
the connection transforms as $A^{f}=g^{-1}Ag+g^{-1}dg$, so that the
nontrivial transformations in terms of the components of the gauge
fields read
\begin{equation}
B^{f}=B+df\qquad,\qquad\xi^{\pm f}=e^{\pm if}\xi^{\pm}\,.
\end{equation}
Therefore, a generic choice of boundary conditions labelled by $\eta$,
can be obtained from the one of periodic boundary conditions ($\eta=0$)
if one chooses $f=\eta\phi$. For this choice, the dynamical fields
that describe the asymptotic structure in \eqref{eq:aphi} then transform
as 
\begin{gather}
\mathcal{J}^{\eta}=\mathcal{J}+\eta\mathcal{T}\quad,\quad\mathcal{P}^{\eta}=\mathcal{P}-4\eta\mathcal{Z}+\frac{k}{\pi}\eta^{2}\,,\nonumber \\
\mathcal{Z}^{\eta}=\mathcal{Z}-\frac{k}{2\pi}\eta\quad,\quad\mathcal{G}^{\pm\eta}=e^{\pm i\eta\phi}\mathcal{G}^{\pm}\,,\label{eq:FieldRed}
\end{gather}
with $\mathcal{T}^{\eta}=\mathcal{T}$. In terms of Fourier modes,
eq. \eqref{eq:FieldRed} reads
\begin{align}
\mathcal{J}_{m}^{\eta}= & \mathcal{J}_{m}+\eta\mathcal{T}_{m}\,,\nonumber \\
\mathcal{P}_{m}^{\eta}= & \mathcal{P}_{m}-4\eta\mathcal{Z}_{m}+2\eta^{2}k\delta_{m,0}\,,\nonumber \\
\mathcal{Z}_{m}^{\eta}= & \mathcal{Z}_{m}-\eta k\delta_{m,0}\,,\label{eq:Automorphism}\\
\mathcal{T}_{m}^{\eta}= & \mathcal{T}_{m}\,,\nonumber \\
\mathcal{G}_{p\pm\eta}^{\eta\pm}= & \mathcal{G}_{p}^{\pm}\,,\nonumber 
\end{align}
and it is then simple to verify that the new ``$\eta$'' generators
fulfill the super-BMS$_{3}$ algebra with $\mathcal{N}=(2,0)$ for
an anyonic choice of boundary conditions. In this sense \eqref{eq:Automorphism}
can be regarded as an automorphism of the asymptotic symmetry algebra
\eqref{eq:alg3} for generic boundary conditions for the fermions
as in \eqref{eq:Anyonic}. Therefore, in particular, the algebras
with Ramond and Neveu-Schwarz boundary conditions become related by
spectral flow. One can then say that a generic choice of boundary
conditions for the fermions can be ``gauged away'' by virtue of the
$U(1)$ automorphism of the super-BMS$_{3}$ algebra with $\mathcal{N}=(2,0)$.
Thus, different theories characterized by inequivalent choices of
boundary conditions, actually coincide after a suitable field redefinition
that is induced through an appropriate $U(1)$ gauge transformation.

\section{Extension of the results \label{Extensions}}

One of the advantages of formulating supergravity in terms of a Chern-Simons
action as in \cite{Achucarro:1987vz}, \cite{Achucarro:1989gm}, \cite{Howe:1995zm},
is that the theory can be extended to include parity odd terms in
the Lagrangian in a straightforward way along the lines of \cite{Giacomini:2006dr}
(see also \cite{Barnich:2014cwa}, \cite{Barnich:2015sca}, \cite{Fuentealba:2015wza}).
The procedure amounts to introduce an additional coupling through
a simple modification of the invariant bilinear form of the gauge
group, as well as further couplings through shifting the magnetic-like
gauge fields by the corresponding ones of electric type. Therefore,
as explained in \cite{Barnich:2014cwa}, \cite{Fuentealba:2015jma},
\cite{Fuentealba:2015wza}, the global charges become suitably modified,
so that the asymptotic symmetry algebra is able to acquire additional
central extensions. It would then be interesting to explicitly construct
this kind of extension for the supergravity theory with $\mathcal{N}=(2,0)$
in \cite{Howe:1995zm}. It is also worth pointing out that this procedure
has recently been applied in \cite{Basu:2017aqn} for the case of
the inhomogeneous (despotic) theory with $\mathcal{N}=(2,0)$, so
that an additional central extension manifestly shows up in the asymptotic
symmetry algebra, as well as through an additional contribution to
the entropy of cosmological spacetimes. 

Another possibility that deserves to be explored is to consider the
extension of our analysis for $\mathcal{N}>2$, since it is natural
to expect that the extended super-BMS$_{3}$ algebra has to be nonlinear,
as it can be directly seen from the flat limit of the superconformal
algebra in two spacetime dimensions. Indeed, nonlinear extensions
of the BMS$_{3}$ algebra have been shown to arise in the context
of hypergravity \cite{Fuentealba:2015jma}, \cite{Fuentealba:2015wza}
as well as for higher spin gravity without cosmological constant in
\cite{Afshar:2013vka}, \cite{Gonzalez:2013oaa}, \cite{Gary:2014ppa},
\cite{Matulich:2014hea}.

As a closing remark, it is worth pointing out that induced representations
of BMS$_{3}$, as well as its extensions that include higher spin
generators have been discussed in \cite{Barnich:2014kra}, \cite{Campoleoni:2015qrh},
\cite{Campoleoni:2016vsh}, \cite{Oblak:2016eij}. In the case of
(higher spin) fermionic fields, this has been done so far for $\mathcal{N}=1$.
It would then be interesting to explore the properties of this sort
of representations of super-BMS$_{3}$ for an extended number of fermionic
generators, as in the case discussed in this work.

\acknowledgments We thank Daniel Grumiller, Wout Merbis, Alfredo
Pérez and David Tempo for enlightening discussions and comments. O.F.
and R.T. wish to thank Daniel Grumiller and the organizers of the
ESI Programme and Workshop ``Quantum Physics and Gravity'' hosted
by the Erwin Schrödinger Institute (ESI), during June of 2017 in Vienna,
for the opportunity of presenting this work. Special thanks to Max
Riegler for letting us know and kindly explaining his interesting
work in collaboration with Rudranil Basu and Stephane Detournay in
\cite{Basu:2017aqn} during the workshop. This research has been partially
supported by Fondecyt grants Nº 3150448, 3170772, 1161311, 1171162.
The Centro de Estudios Científicos (CECs) is funded by the Chilean
Government through the Centers of Excellence Base Financing Program
of Conicyt.

\appendix
%dummy comment inserted by tex2lyx to ensure that this paragraph is not empty%dummy comment inserted by tex2lyx to ensure that this paragraph is not empty%dummy comment inserted by tex2lyx to ensure that this paragraph is not empty

\section{Conventions \label{Conventions}}

We choose the orientation to be such that $\varepsilon_{012}=1$.
The Minkowski metric $\eta_{ab}$ is assumed to be non-diagonal, so
that $\eta_{01}=\eta_{10}=\eta_{22}=1$, while the remaining components
vanish. In the spinorial representation, the generators of $SO(2,1)$
are given by $\left(J_{a}\right)_{\beta}^{\alpha}=\frac{1}{2}\left(\Gamma_{a}\right)_{\beta}^{\alpha}$,
where the matrices $\Gamma_{a}$ fulfill the Clifford algebra, $\left\{ \Gamma_{a},\Gamma_{b}\right\} =2\eta_{ab}$.
They are chosen in terms of the Pauli matrices $\sigma_{i}$, so that
\begin{equation}
\Gamma_{0}=\frac{1}{\sqrt{2}}\left(\sigma_{1}+i\sigma_{2}\right)\quad,\quad\Gamma_{1}=\frac{1}{\sqrt{2}}\left(\sigma_{1}-i\sigma_{2}\right)\quad,\quad\Gamma_{2}=\sigma_{3}\,,
\end{equation}
with
\begin{equation}
\sigma_{1}=\left(\begin{array}{cc}
0 & 1\\
1 & 0
\end{array}\right)\quad,\quad\sigma_{2}=\left(\begin{array}{cc}
0 & -i\\
i & 0
\end{array}\right)\quad,\quad\sigma_{3}=\left(\begin{array}{cc}
1 & 0\\
0 & -1
\end{array}\right)\,.
\end{equation}
Spinors $\psi^{\alpha}$ are labeled according to $\alpha=+,-$, and
we define the Majorana conjugate as $\bar{\psi}_{\alpha}=\psi^{\beta}C_{\beta\alpha}$,
where $C_{\alpha\beta}$ stands for the charge conjugation matrix,
given by
\begin{equation}
C_{\alpha\beta}=C^{\alpha\beta}=\left(\begin{array}{cc}
0 & -1\\
1 & 0
\end{array}\right)\,,
\end{equation}
so that $C^{\alpha\beta}$ is the inverse. The charge conjugation
matrix then fulfills $C^{T}=-C$, and $\left(C\Gamma_{a}\right)^{T}=C\Gamma_{a}$.

In the fundamental representation of $SL(2,\mathbb{R})\times U(1)$,
the generators of $SL(2,R)$ and the $U(1)$ generator are given by
$L_{m}$, with $m=-1,0,1$, and $T$, respectively. They are chosen
as
\begin{equation}
L_{-1}=\left(\begin{array}{cc}
0 & -1\\
0 & 0
\end{array}\right)\quad;\quad L_{0}=\left(\begin{array}{cc}
\frac{1}{2} & 0\\
0 & -\frac{1}{2}
\end{array}\right)\quad;\quad L_{1}=\left(\begin{array}{cc}
0 & 0\\
1 & 0
\end{array}\right)\quad;\quad T=\left(\begin{array}{cc}
i & 0\\
0 & i
\end{array}\right)\,.
\end{equation}
Therefore, the nonvanishing components of the trace of quadratic products
of them become $\text{tr}\left(L_{1}L_{-1}\right)=-1$, $\text{tr}\left(L_{0}^{2}\right)=\frac{1}{2}$
and $\text{tr}\left(T^{2}\right)=-2$.

\end{document}